\documentclass[12pt,letterpaper]{article}
\usepackage[margin=1in]{geometry}
\usepackage[utf8]{inputenc}
\RequirePackage{amsthm,amsmath,amsfonts,amssymb}
\usepackage{natbib}
\PassOptionsToPackage{hyphens}{url}
\RequirePackage{graphicx}
\usepackage{mathabx}
\usepackage{xr} 
\usepackage{amsmath}
\usepackage{amssymb}
\usepackage[normalem]{ulem}
\usepackage[]{natbib}
\usepackage[dvips]{epsfig}
\usepackage{dcolumn}
\usepackage{enumerate}
\usepackage{hhline}
\usepackage{dsfont}
\usepackage{afterpage}
\usepackage{arydshln}
\usepackage{graphicx}
\usepackage{color}
\usepackage{pifont}
\usepackage{setspace}
\usepackage{booktabs}
\usepackage{placeins}
\newcommand{\cmark}{\ding{51}}
\newcommand{\xmark}{\ding{55}}
\newcommand{\greencheck}{{\color{pinegreen}\cmark}}
\newcommand{\redcross}{{\color{red}\xmark}}

\usepackage{amsmath, amsthm, amssymb}
\usepackage{tikz}
\usepackage{enumitem}
\usetikzlibrary{arrows.meta}

\newtheorem{proposition}{Proposition}[section]

\newtheorem{theorem}[proposition]{Theorem}
\newtheorem{corollary}[proposition]{Corollary}

\theoremstyle{definition}

\theoremstyle{plain}
\newtheorem{assumption}{Assumption}

\usepackage{rotating}
\usepackage{breakurl} 
\usepackage{xr}
\usepackage{xr-hyper}
\RequirePackage[colorlinks,citecolor=blue,urlcolor=blue,breaklinks]{hyperref}
\usepackage{hyperref}
\usepackage[percent]{overpic}
\usepackage{subfig}
\usepackage[]{caption}
\usepackage{algorithmicx}
\usepackage[noend]{algpseudocode}
\usepackage{algorithm}
\usepackage{diagbox}
\usepackage{graphicx}
\usepackage{wrapfig}
\usepackage{lscape}
\input epsf
\usepackage{fontenc}
\usepackage{bm}
\usepackage{multirow}
\usepackage{eurosym}
\epsfverbosetrue

\DeclareMathAlphabet{\mathpzc}{OT1}{pzc}{m}{it}
\DeclareFontFamily{OT1}{pzc}{}
\DeclareFontShape{OT1}{pzc}{m}{it}{<-> s * [1.200] pzcmi7t}{}

\def \dsP {\text{$\mathds{P}$}}
\def \dsE {\text{$\mathds{E}$}}
\def \dsR {\text{$\mathds{R}$}}

\DeclareMathOperator{\pen}{\scriptstyle{nonlin}}
\DeclareMathOperator{\unpen}{\scriptstyle{lin}}
\DeclareMathOperator{\notsel}{\scriptstyle{in}}

\DeclareMathOperator{\ND}{N}

\DeclareMathOperator{\GaD}{Ga}

\DeclareMathOperator{\IGD}{IG}

\DeclareMathOperator{\ExpD}{Exp}

 \DeclareMathOperator{\BerD}{\mathpzc{Be}}
  \newcommand{\BetaD}{\mathpzc{Beta}}

    \def \mD {\text{\boldmath$D$}}

\definecolor{parisgreen}{rgb}{0.31, 0.78, 0.47}
\definecolor{pinegreen}{rgb}{0.0, 0.47, 0.44}
\definecolor{midnightblue}{rgb}{0.1, 0.1, 0.44}
\definecolor{orangered}{rgb}{1.0, 0.27, 0.0}
\definecolor{burntorange}{rgb}{0.8, 0.33, 0.0}
\definecolor{amethyst}{rgb}{0.6, 0.4, 0.8}

\theoremstyle{remark}

\definecolor{redP}{HTML}{F64A8A}

\newcommand{\modname}{{$\mathsf{QDReSS}$}}
\newcommand{\modnamesp}{{$\mathsf{QDReSS}$ }}

\begin{document}
\setlength{\abovedisplayskip}{0.15cm}
\setlength{\belowdisplayskip}{0.15cm}
\pagestyle{empty}
\newgeometry{margin=1in,top=0.5in}
\begin{titlepage}
\pagestyle{empty}
\vspace{-30pt}
\title{
\singlespacing \bf Bayesian Effect Selection for Additive Quantile Regression  with an Application to Air Pollution Thresholds}
\author{Nadja Klein$^1$, Aaron Wei Qi Lee$^1$ and Jorge Mateu$^{2}$\\
   \normalsize $^1$Scientific Computing Center, Karlsruhe Institute of Technology, Germany\\
\normalsize $^2$Department of Mathematics, University Jaume I, 12071 Castellon, Spain}
\date{}
\maketitle
\vspace{-30pt}

\begin{abstract}
 \singlespacing
\vspace{-10pt}
Air pollution regulatory limits are typically defined in terms of exceedances of concentration thresholds. Such exceedances are naturally related to the upper conditional quantiles of the pollutant distribution and are therefore of direct relevance for assessing severe pollution events. At the same time, it is important to determine not only whether a covariate affects air pollution but also whether this effect is linear, nonlinear, or both. We address these issues by developing an interpretable Bayesian effect selection approach for additive quantile regression. While commonly used mixed model representations (MMRs) of penalized splines allow for flexible nonlinear effects, they do not provide a meaningful separation of linear and nonlinear effect components. We therefore employ a Demmler-Reinsch basis expansion, which yields an orthogonal decomposition of each additive effect into linear and nonlinear parts and show theoretically that both effect components can be estimated consistently. To facilitate data-driven model building, we propose a Bayesian effect selection approach with separate normal beta prime spike and slab priors on the scalar importance parameters associated with the linear and nonlinear components and implement efficient posterior estimation using a  Gibbs sampler. Through simulation studies, we demonstrate robustness to the misspecification induced by the employed asymmetric Laplace working likelihood and show superior performance relative to the MMR. We finally apply the proposed method to air pollution data in Madrid, Spain. The results highlight the added value of flexibly modeling extreme nitrogen dioxide (NO$_2$) concentrations and reveal that threshold-relevant pollution levels are driven differently by climatological variables and traffic-related spatial structure. These findings underline the need for advanced statistical models that support short-term decision-making and help local authorities mitigate, or potentially prevent, exceedances of NO$_2$ concentration limits.
\end{abstract}
\vspace{-15pt}
\singlespacing
\noindent
{\bf Keywords}: Air quality, asymmetric Laplace distribution, Demmler-Reinsch basis, distributional regression, effect decomposition, NO$_2$ and O$_3$, spike and slab prior\vspace{0.5cm}\\
 \noindent
{\footnotesize{{\bf Corresponding author}: Nadja Klein gratefully acknowledges support  by the German research foundation (DFG) through the  Emmy Noether grant KL 3037/1-1. Jorge Mateu is partially supported by 
{grant PID2022-141555OB-I00 from Ministry of Science and Innovation}. The authors thank Nicolas Bianco for proofreading the simulation study. }}

\end{titlepage}
\restoregeometry
\newpage
\pagestyle{plain}

\section{Introduction}\label{sec:intro}
Air quality is deteriorating globally at an alarming rate due to increasing industrialization
and urbanization~\citep{Ryu2019}. According to the European Environment Agency \citep{EEA2022a}, air pollution is the single largest environmental health risk in Europe, and in many countries worldwide. It is a major cause of adverse health effects, as air pollution causes and aggravates respiratory and cardiovascular diseases. Heart disease and stroke are the most common causes of premature deaths attributed to air pollution, followed by lung disease and lung cancer. In 2022, air pollution led to a significant number of premature deaths in the 27 EU Member States (EU-27). Indeed, following \cite{EEA2022b}, we can underline these three critical concerns: (a) exposure to concentrations of fine particulate matter above the 2021 World Health Organization guideline level resulted in 238,000 premature deaths, (b) exposure to nitrogen dioxide above the respective guideline level led to 49,000 premature deaths, and (c) acute exposure to ozone caused 24,000 premature deaths. The Urban NO$_2$ Atlas \citep{NO2Atlas} provides city fact sheets to help design effective air quality measures with the objective of reducing NO$_2$ concentration within European cities, noting that many European cities still regularly exceed the current EU limits for NO$_2$. It also identifies the main sources of NO$_2$ pollution in each city, helping policymakers design target measures and actions against them. Some of the outlined sources are motorized transport for traveling (private cars and heavy goods vehicles) that produce nitric oxide (NO) and CO, and more importantly O$_3$ that reacts with NO to produce NO$_2$. Furthermore, NO$_2$ and NO interact with water, oxygen and other chemicals in the atmosphere to form acid rain.

{NO$_2$} is one of a group of gases called nitrogen oxides (NO$_\text{x}$). Although all of these gases are harmful to human health and the environment, NO$_2$ is of greater concern~\citep{Val2011,Ach2019} and is mainly emitted into the air through burning of fuel. NO$_2$ is generated by emissions from cars, trucks and buses, power plants, and off-road equipment~\citep{Per2001,Lee2014,Cat2016} and is a primary pollutant. Indeed, as reported by the European Environment Agency~\citep{EEA2022a}, road transport is the main contributor to NO$_2$ pollution in the EU, ahead of the energy, commercial, institutional, and household sectors. On the other {hand}, the O$_3$ molecule{s} {are} harmful to air quality outside of the ozone layer. O$_3$ is a so-called secondary pollutant that is not emitted directly from a source (such as vehicles or power plants).
O$_3$ can be ``good'' or ``bad'' for health and the environment depending on where it is found in the atmosphere. Stratospheric ozone is ``good'' because it protects the living organisms from ultraviolet radiation from the sun. Ground-level ozone is a colorless and highly irritating gas that forms just above the earth's surface. It is ``bad'' because it can trigger a variety of human health problems, particularly for children, the elderly, and people of all ages who have lung diseases such as asthma~\citep{Val2011}.
 Breathing air with a high concentration of NO$_2$ can irritate the airways in the human respiratory system and can harm our health.
As the global population becomes more health conscious, various studies have been conducted to determine the effect of NO$_2$ concentration on human health. High concentrations of NO$_2$  in urban areas can cause bronchial and lung cancer and have severe effects on asthmatic patients~\citep{Ach2019}.

Social medical costs due to NO$_2$ pollution are certainly high. To reduce these social costs, many countries regulate NO$_2$ concentration levels using environmental policies aimed at reducing NO$_2$ \citep{Val2011,Bor2018,Ryu2019}. However, to establish effective environmental regulations that reduce the impact of NO$_2$, accurate information on the nature and the way several sources of air pollution interact with and/or produce NO$_2$ is crucial. Modeling air pollution at both global and local scales enables consistent comparisons of relations between air pollution and health. For example, since NO$_2$ is highly traffic-related and localized, its local understanding is needed to assess personal outdoor exposure, in particular in areas close to pollutant sources such as primary roads. Finding the hidden relationship of NO$_2$ and other pollutants, such as CO or O$_3$ is rather complex due to hidden chemical processes, and it is at the core of our motivation here.

Statistical models for air pollution analysis are continuously being developed. Most classical approaches are based on regression techniques through linear-based relations~\citep{Men2015,Hat2017,Ant2018,Lua2020}. However, forecasting pollutant concentrations is difficult
because air quality data are often nonlinear, nonstationary, seasonally varying, and usually are a result of complex chemical reactions \citep{Nag2023,Panja2026}. Classical time series models, particularly ARIMA or SARIMA, have long been used for pollutant forecasting because of their interpretability, modest data requirements, and the ability to capture short-term temporal
dependence \citep{Hyn2018}. However, ARIMA-type models are fundamentally linear and primarily temporal; they struggle to represent nonlinear pollutant
dynamics often present in real-world air quality records \citep{Pak2018}.

Alternatively, tree-based ensembles (e.g.,~random forests) and neural network techniques \\ ~\citep{Ryu2019,Pry2000} have been considered to investigate whether more flexible models can better capture nonlinear relationships between predictors and a response variable, such as NO$_2$. These methods are data-driven and impose fewer distributional assumptions, making them suitable for nonlinear and noisy environmental data. However, these models may be sensitive to  outliers, do not naturally provide structured learning or calibrated probabilistic forecasts, limit interpretability, and their scalability can become restrictive in dense monitoring networks.

Regression-based methods fit one model to the
entire range of each predictor, while ensemble tree-based methods build on subsets of data and sub-ranges of predictors. However, despite the increased flexibility, such models lack the ability to understand and interpret which predictors affect air pollution and in what functional form (e.g.,~linear or nonlinear). Both approaches, in addition, come with the drawback of modeling  the expected NO$_2$ concentration only, rather than allowing to directly regress on exceedances over certain thresholds  of the entire NO$_2$ distributions.  
Such thresholds are imposed by authorities to decide on and classify the severity of a particular event, and they play a key role in decision making.  From a statistical point of view, these thresholds induce quantiles, and this is why we here propose the natural route (albeit not much treated) of modeling NO$_2$ exceedances over certain thresholds using quantile regression. In addition to allowing the predictor effects to differ for distinct quantiles, we develop quantile regression that allows for general function selection and effect decomposition into respective linear and nonlinear parts.
Being able to select the form in which a predictor enters the model, and whether a covariate effect is linear or nonlinear, represents a necessary step forward in the modeling strategy when it comes to targeting the right relationships between predictors and responses in air pollution problems. This type of data is inherently complex, and any effort to provide data-driven effect decomposition, will result in a more accurate understanding of the problem.

Let $\mathcal{D}=\{(y_i,x_i)\}_{i=1}^n$ denote the data with $n$ pairs of independent observations $y_i$ on a continuous response variable $Y\in\mathcal{Y}\subseteq\mathds{R}$, and $x_i$ the corresponding vector of covariates. Estimation of the conditional $\tau$-th quantile of interest, $\tau\in(0,1)$, was  originally proposed by \citet{KoeBas1978} in a  linear quantile regression (QR) model. It is based on solving 
\begin{equation}\label{eq:awad}
    \min_{\beta_0\in\dsR,\;\beta\in\dsR^p}\sum_{i=1}^n \rho_\tau(y_i-\beta_0-x_i^\top\beta),
\end{equation}
where $\rho_{\tau}(u)=u(\tau-\mathds{1}_{\lbrace u<0\rbrace})$ for $u\in\dsR$ is the so-called piecewise linear ``check function'', and $\beta_0\in\dsR,\;\beta\in\dsR^p$ are the intercept and regression coefficients. This approach is fully non-parametric as it does not make any specific choice about an error distribution. 

To be able to handle nonlinear or more general functional relationships between response and covariates on conditional quantiles directly, structured additive quantile (STAQ) regression models have been suggested in the literature~\citep{FenKneHot2011,YueRue2011,WalKneLanYue2012,FasWooZafNedGou2021}. These are based on the general idea of structured additive models for mean regression~\citep[STAR;][]{Woo2017}.

However, when it comes to general effect decomposition and selection in STAQ models, the literature becomes much scarcer. Indeed, effect selection for quantile regression has so far mostly been done in the linear case. A prominent example is the Least Absolute Shrinkage and Selection Operator  \citep[LASSO;][]{LiZhu2008}, which imposes an $L_1$-penalty on the regression coefficients to achieve simultaneous estimation and variable selection. Building upon this idea, several variants have been proposed to improve selection consistency, such as  adaptive LASSO \citep{WuLiu2009}, and (non-convex) penalties such as the Smoothly Clipped Absolute Deviation  penalty \citep{WuLiu2009}. In the non-parametric regression context, the COmponent Selection and Smoothing Operator \citep[COSSO;][]{LinZha2006} unifies model selection and function estimation by penalizing the functional components of an additive model. COSSO effectively selects entire component functions, and while the authors do mention that effect decomposition into linear and nonlinear parts could be possible, its implementation does not explicitly do so. The extension of COSSO to quantile regression \citep{LinBonZhaZou2013} inherits this limitation.

In the Bayesian quantile regression framework, variable selection has been proposed using different  prior distributions, such as Bayesian versions of the LASSO \citep{Alh2015},  group lasso or elastic net   \citep{LiXiLin2010}, and horseshoe priors \citep{KohSze2024}. Alternative methods such as stochastic search variable selection make use of a variation of a spike and slab prior \citep{YuCheReeDun2013}.  

To our knowledge, the only existing framework that explicitly integrates effect decomposition with variable selection in quantile regression, while simultaneously determining whether a covariate enters the model linearly or nonlinearly, is the statistical boosting framework proposed by \cite{FenKneHot2011}. However, statistical boosting comes with its own limitations, such as biased estimators due to early stopping or costly to achieve standard errors. Furthermore, the authors rely on the mixed model representation (MMR) of \citet{FahKneLan2004}, which does not provide an orthogonal effect decomposition and yields thus inconsistent estimation of respective linear and nonlinear effect parts \citep{BacKle2024}. Table \ref{tab:methods_literature} gives an overview of existing approaches and their features. 

\begin{table}[!t]
  \centering
  \resizebox{1\textwidth}{!}{
    \begin{tabular}{cccccccccc}
    \toprule
        \bf{Bayesian} &   & &   &  & & & & & \\
    \toprule
&  \cite{LiXiLin2010} & & \cite{YuCheReeDun2013} & & \cite{Alh2015} & & \cite{KohSze2024} & & \modnamesp (Ours) \\
        \cmidrule{2-2}\cmidrule{4-4}\cmidrule{6-6}\cmidrule{8-8}\cmidrule{10-10}
        Nonlinear Effect &   \redcross & & \redcross & & \redcross & & \redcross & & \greencheck \\
        Effect Decomposition & \redcross & & \redcross & & \redcross & & \redcross & & \greencheck \\
        \textsf{R} Package &   \redcross & & \greencheck ( \texttt{SSVSquantreg}) & & \greencheck (\texttt{Brq}) & & \greencheck ( \textsf{GitHub}) & & \greencheck ( \textsf{GitHub}) \\
    \bottomrule
    \toprule
    \bf{Non-Bayesian} &   & &   &  & & & & & \\
    \toprule
&  \cite{LiZhu2008} & & \cite{WuLiu2009} & & \cite{FenKneHot2011} & & \cite{LinBonZhaZou2013} & & \cite{MugTorEilSciAtt2021} \\
        \cmidrule{2-2}\cmidrule{4-4}\cmidrule{6-6}\cmidrule{8-8}\cmidrule{10-10}
        Nonlinear Effect &  \redcross & & \redcross & & \greencheck & & \greencheck & & \greencheck \\
        Effect Decomposition & \redcross & & \redcross & & \greencheck & & \redcross & & \redcross \\
        \textsf{R} Package &  \redcross  & & \redcross & & \greencheck ( \texttt{mboost}) & & \greencheck ( \texttt{cosso}) & & \greencheck ( \texttt{quantregGrowth})  \\
    \bottomrule
    \end{tabular}
  }
  \vspace{0.25em}
  \caption{Overview of existing methods for variable selection in quantile regression, with Bayesian approaches listed in the top rows and non-Bayesian approaches in the bottom rows. The table entries indicate whether a method allows for nonlinear effects, performs variable selection, enables effect decomposition, and whether an associated R package is available. A green check mark (\greencheck) indicates that the feature is present, while a red cross (\redcross) indicates that it is absent. } \label{tab:methods_literature}
\end{table}

To fill existing gaps and motivated by our application to air pollution, we develop a general Bayesian effect selection framework for STAQ models. Rather than employing the commonly suggested MMR, we consider the recently proposed Demmler-Reinsch (DR) basis expansion of \citet{BacKle2024}, which allows us to consistently estimate the linear and nonlinear effect components through an orthogonal decomposition. As a theoretical contribution, we show that this decomposition also yields consistent estimators of the quantile-specific linear and nonlinear effect components in STAQ. We then enable data-driven selection of quantile-specific effects and their functional forms by combining the DR  basis expansion with the normal beta prime spike and slab (NBPSS) prior of \citet{KleCarKneLanWag2021}.  We denote our approach \modnamesp and implement posterior estimation based on efficient Markov chain Monte Carlo (MCMC) simulations.  All steps can be realized by Gibbs updates, allowing for fast sampling. 

In summary, the combination of DR bases with the NBPSS prior yields a coherent,  
 computationally efficient and consistent framework for effect selection in additive 
quantile regression. 
The DR basis guarantees empirical orthogonality between linear and nonlinear 
components (as opposed to the commonly employed MMR), ensuring that selection decisions are identifiable and 
interpretable. 
At the same time, the hierarchical spike and slab prior enforces group-level 
shrinkage. 
This construction enables the proposed model to distinguish between linear, 
nonlinear, and null effects in a principled Bayesian manner, while retaining 
the flexibility needed for complex applications.

Bayesian variable selection based on spike and slab priors was {popularized} by~\citet{MitBea1988} for linear mean regression and  has since been further developed significantly in several contexts~\citep[for instance,][]{GeoMcC1997}; see also \citet{ClyGeo2004,OHaSil2009} for some overviews. We believe our  approach makes a necessary contribution and specifically enables  three essential features simultaneously: to (i) study  quantile-specific  covariate effects, (ii) allow these covariates to be of general functional form (e.g.,~nonlinear) using additive predictors, (iii) decide whether an effect should be included linearly, nonlinearly or not at all in the  relevant threshold quantiles. 

The remainder of the paper is structured as follows. 
Section~\ref{sec:data} introduces the Madrid air pollution data and formulates the research questions that motivate the proposed methodology. Section~\ref{sec:effselQR} develops consistent effect decomposition and selection for additive quantile regression. Section~\ref{sec:bayes}  introduces Bayesian estimation for the proposed model, including the NBPSS prior specification and the Gibbs sampling scheme. Section~\ref{sec:results} presents the results of the application to NO$_2$ concentrations in Madrid and quantifies the added value of modeling extreme NO$_2$ concentrations, and how thresholds that induce conditional quantiles can highlight important modeling differences. These results underpin the need of
enhanced statistical models to support short-term decisions and enable local authorities to mitigate or even prevent exceedances of NO$_2$ concentration limits. Section~\ref{sec:conclusion} concludes and discusses the limitations of our study. In the Supplementary Material we provide a detailed simulation study that demonstrates that \modnamesp is capable of consistently selecting and estimating quantile-specific linear and nonlinear effects along with accurate uncertainty quantification, and provide further computational and theoretical details, as well as MCMC diagnostics for the real data analysis.

\section{The data and research questions}\label{sec:data}

\subsection{Air pollution, NO$_2$ and O$_3$. Some motivating basics}\label{subsec:motivation}

Air pollution in urban areas is mainly due to the intense use of motorized transport for traveling, in particular private cars and heavy goods vehicles. This is a priority issue for transportation planners and public authorities, given the harmful effects of pollution to human health and the environment~\citep{Berg2013}. 
NO$_2$ along with  nitric oxide (NO) reacts with other chemicals in the air, and both of these are harmful when inhaled due to effects on the respiratory system. NO$_2$ and NO interact with water, oxygen and other chemicals in the atmosphere to form acid rain. Acid rain harms sensitive ecosystems such as lakes and forests. Breathing air with a high concentration of NO$_2$ can irritate airways in the human respiratory system, and exposures over short periods can aggravate respiratory diseases. People with asthma, as well as children and the elderly, are generally at greater risk for  the health effects of NO$_2$. A related air pollutant is 
ground{-}level O$_3$ which is not emitted directly into the air, but formed through chemical reactions between NO$_\text{x}$ and volatile organic compounds (VOCs). Exposure to elevated ozone levels can adversely harm human health, particularly among individuals with certain genetic predispositions and those with lower intake of certain nutrients such as vitamins C and E. O$_3$ exposure can also exacerbate respiratory conditions. Thus, clear representatives of air pollution in urban areas are the  photochemical oxidants {O$_3$} and NO$_2$~\citep{Who2014}.

The formation of {O$_3$} depends on the intensity of solar radiation, the
absolute concentrations of NO$_\text{x}$ and VOC, and the ratio of NO$_\text{x}$ (NO and NO$_2$) to VOC~\citep{Val2011}. NO is converted to NO$_2$ via a reaction with O$_3$, However, this phenomenon is not well understood so far. The concentration of photochemical oxidants can be decreased by controlling their precursors: nitrogen oxides NO$_\text{x}$  and VOCs. However, the efficiency of emission control also depends on the relationship between primary and secondary pollutants, as well as the ambient meteorological conditions. 
Clearly, and owing to the chemical coupling of O$_3$ and NO$_\text{x}$, the levels of O$_3$ and NO$_2$ are inherently linked in a rather complex form, so complex that the response to reduction in the emission of NO$_\text{x}$ is remarkably nonlinear~\citep{Val2011,Bor2018,Ryu2019} and any resultant reduction in the level of NO$_2$ is invariably accompanied by an increase in the level of O$_3$. In addition, changes in the level of O$_3$ on a global scale lead to an increasing background which influences local O$_3$ and NO$_2$ levels and the effectiveness of local emission controls.

It is therefore necessary to obtain a thorough understanding of the cross relationship among O$_3$, traffic flow and NO$_2$ under various atmospheric conditions to improve the understanding of the chemical coupling among them. The nonlinear mechanisms of the inter-dependencies between O$_3$ and NO$_2$ in combination with other pollutants and atmospheric conditions are still not well understood and need further study. These nonlinear effects are in line with thresholds that authorities impose on the records of NO$_2$; environmental pollution alarms are placed upon three different thresholds, 60\% (moderate), 80\% (large) and 90\% (extreme) \citep{EEA2022b}.

\subsection{{The data}}\label{subsec:datades}

The Surveillance System of the City of Madrid (Spain) keeps a web portal with open data from many sources and environmental problems (see \burl{https://datos.madrid.es/portal/site/egob}). In particular, this system collects basic information through a number of stations  for atmospheric
surveillance. We focus here on one weather and one pollution station located in downtown Madrid as they are placed in a strategic part of the city that represents one of the peak locations for air pollution and traffic congestion. 
We have daily data from 1 January 2016 to 31 December 2019 on the three air pollutants NO$_2$ ($\mathit{no2}$), O$_3$ ($\mathit{o3}$) and CO (carbon monoxide, $\mathit{co}$), {together with a number of climatological variables}. For the latter, we have the daily average precipitation ($\mathit{prec}$), temperature ($\mathit{temp}$), average wind speed {($\mathit{vel}$)}, wind gust speed ($\mathit{racha}$), maximum pressure ($\mathit{pres\_max}$), and minimum pressure ($\mathit{pres\_min}$). In addition, we have information on 
traffic flow {($\mathit{traffic}$)} averaged per day and per street, together with the maximum and minimum per day of 800 streets surrounding the measuring station.
Traffic and pollution data provided by the city council of Madrid is available at \burl{https://datos.madrid.es/portal/site/egob}. 

To account for systematic differences across calendar years, we include year-specific indicator variables. 
Specifically, the period 2016--2019 is represented by three dummy variables 
$\mathit{year}_1$, $\mathit{year}_2$ and $\mathit{year}_3$, corresponding respectively to the years 2017, 2018 and 2019, while 2016 is used as the reference category. 
These variables allow the model to capture inter-annual shifts in NO$_2$ levels that may be due to changes in traffic patterns, emissions, regulation, or other year-level factors not fully explained by the observed daily meteorological and pollution covariates. We note that patterns can be different in consecutive years and this is better reflected by considering indicator functions rather that smooth functions that assume certain degree of continuity, particularly given that we observe only four distinct time points.

As noted above, and due to complex chemical reactions, NO$_2$ is cross-linked with O$_3$ and other pollutants through nonlinear forms in such a way that any increase or reduction in the level of O$_3$ affects the level of NO$_2$. In this line, we consider the latter as a response variable in our model to fully address these types of relationships for further control strategies.

Hence, for a specific quantile $\tau$, a flexible predictor specification $\eta_\tau$ considered later is \begin{equation}\label{eq:predapp}\begin{aligned}
\eta_{\tau}&=\beta_{\tau,0}+\sum_{k=1}^3 \beta_{\tau, k, \notsel} \mathit{year}_k + f_{\tau,1}(\mathit{co}) + f_{\tau,2}(\mathit{o3}) + f_{\tau,3}(\mathit{prec}) + f_{\tau,4}(\mathit{temp}) + f_{\tau,5}(\mathit{vel})  \\
&\quad\quad\quad + f_{\tau,6}(\mathit{racha})+ f_{\tau,7}(\mathit{pres\_max})+ f_{\tau,8}(\mathit{pres\_min}) + f_{\tau,9}(\mathit{traffic}),
\end{aligned}\end{equation}
where $\beta_{\tau,0}$ and $\beta_{\tau, k, \notsel}$, $k=1,\ldots,3$ are the overall intercept and year-specific coefficients, respectively, and the effects $f_{\tau,j}$, $j=1,\ldots,9$ represent nonlinear effects of the continuous covariates.

\subsection{Descriptive statistics and preliminary analyses}
Table \ref{tab:data} shows a description and summary statistics of the continuous covariates (before standardization to $[0,1]$, which we consider later to facilitate effect selection). The year  2016--2019 is coded as 0/1 dummy variables with 2016 as a reference category.

\begin{table}[htbp] \renewcommand{\arraystretch}{1}
\centering\begin{tabular}{c|cccc}
  \hline\hline
Variable & Description & Mean & Std & Min/Max \\ 
  \hline
$\mathit{co}$ & carbon monoxide & 0.41 & 0.25 & 0.00/2.40 \\ 
  $\mathit{o3}$ & ozone & 39.57 & 20.29 & 0.00/89.00 \\ 
  $\mathit{prec}$ & precipitation & 1.01 & 3.35 & 0.00/28.50 \\ 
  $\mathit{temp}$ & average~temperature & 16.40 & 7.90 & 2.10/32.90 \\ 
  $\mathit{vel}$ & average ~wind speed & 1.80 & 1.00 & 0.00/6.40 \\ 
  $\mathit{racha}$ & wind gust speed & 8.94 & 3.28 & 1.90/26.10 \\ 
  $\mathit{pres\_max}$ & maximum~pressure & 943.20 & 5.58 & 922.00/967.30 \\ 
  $\mathit{pres\_min}$ & minimum~pressure & 938.75 & 6.17 & 915.00/957.40 \\ 
  $\mathit{traffic}$ & average~traffic flow & 776.22 & 147.36 & 249.63/1210.22 \\ 
   \hline\hline
\end{tabular}

\caption{Description and summary statistics of continuous covariates (before standardization to $[0,1]$ in the data set. The average traffic is measured as daily average from  approximately 800 segments of traffic flow around the Carmen station. In addition, we have the year from 2016--2019 coded as 0/1 dummy variables with 2016 as a reference category.}\label{tab:data}
\end{table}

\begin{figure}[htbp]
\centering\includegraphics[width=0.48\textwidth,angle=0]{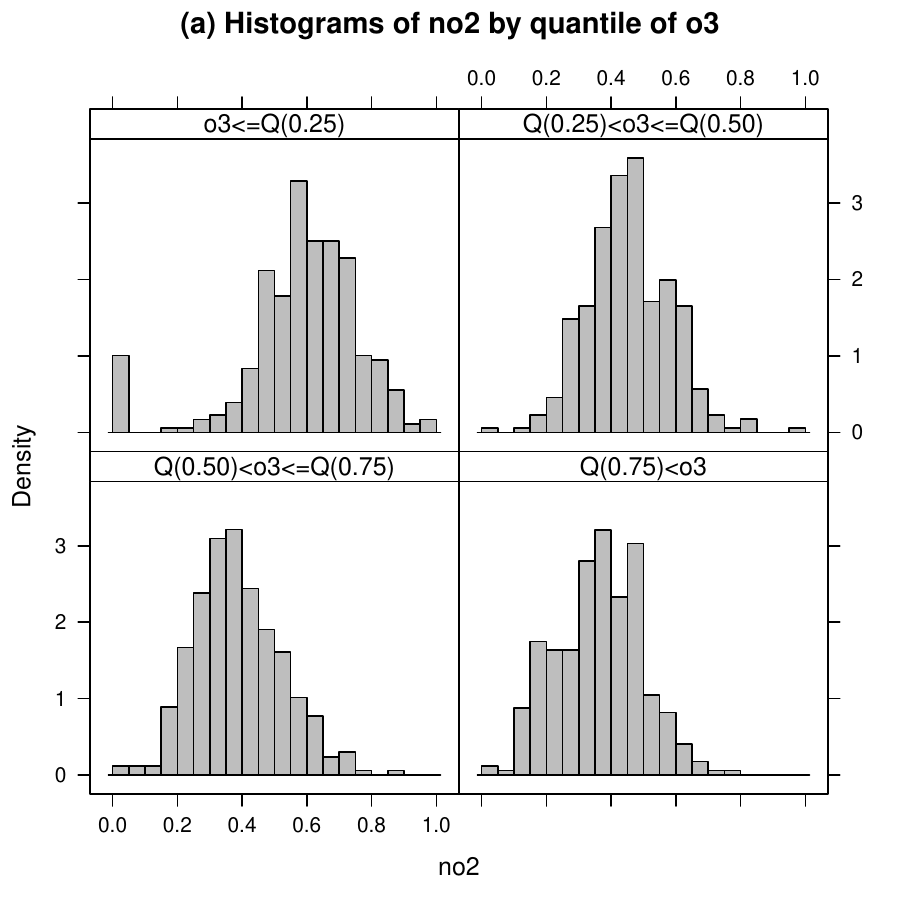}
\centering\includegraphics[width=0.48\textwidth,angle=0]{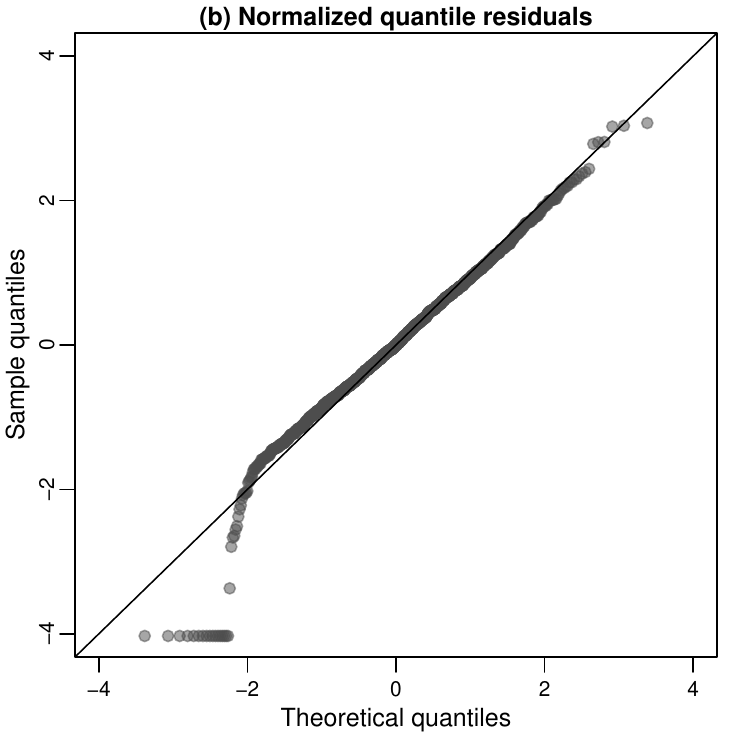}\\
\centering\includegraphics[width=0.48\textwidth,angle=0]{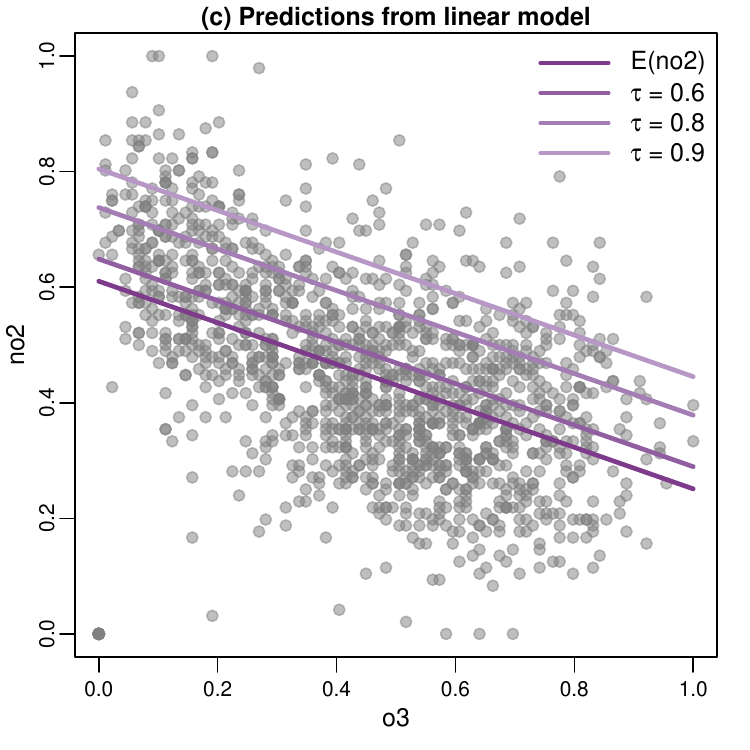}
\centering\includegraphics[width=0.48\textwidth,angle=0]{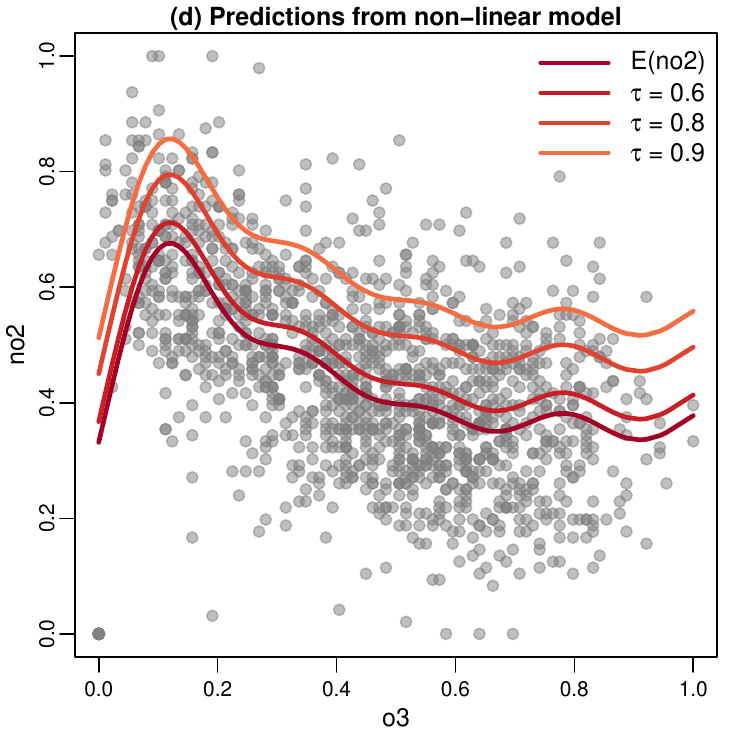}
\caption{Results from univariate preliminary analyses with $\mathit{o3}$ as predictor. Panel (a) shows histograms of the response $\mathit{no2}$ according to the $\mathit{o3}$-quartiles, i.e.~where $Q(\tau)$, $\tau\in\lbrace 0.25,0.50,0.75\rbrace$ is the corresponding quartile of the empirical $\mathit{o3}$ distribution. Panel (b) shows normalized quantiles residuals from the univariate Gaussian regression model. Panels (c) and (d) show the predicted expectation $\dsE(\mathit{no2})$ as well as threshold quantiles $\tau\in\lbrace  0.6,0.8,0.9\rbrace$ obtained from a linear and nonlinear Gaussian model, respectively.}
\label{fig:prelim:no2:o3}
\end{figure}

\begin{figure}[htbp]
\centering\includegraphics[width=0.48\textwidth,angle=0]{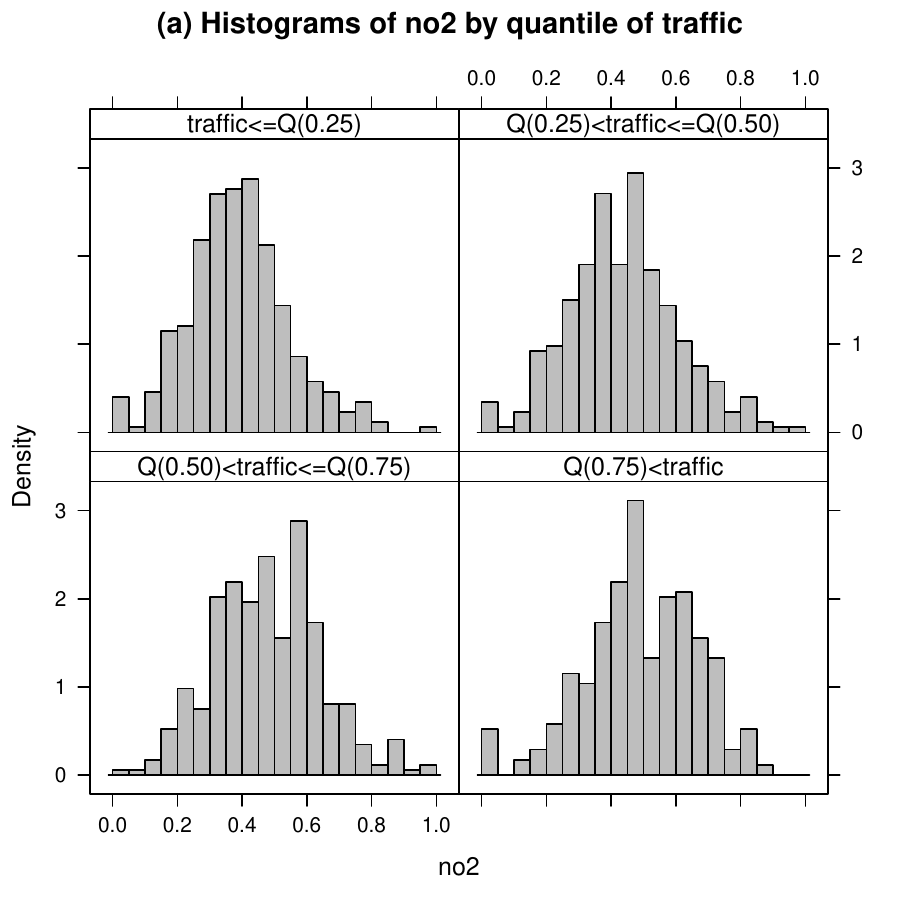}
\centering\includegraphics[width=0.48\textwidth,angle=0]{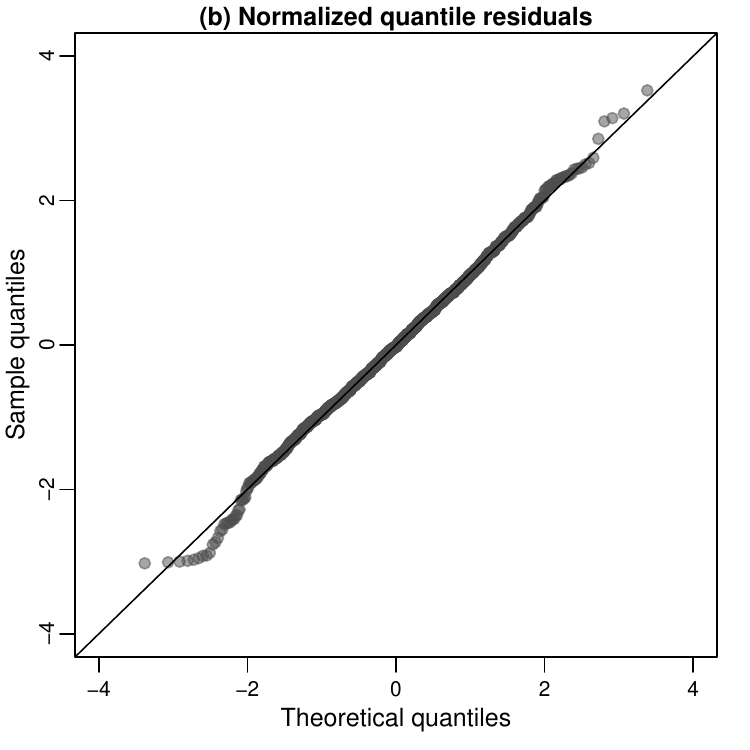}\\
\centering\includegraphics[width=0.48\textwidth,angle=0]{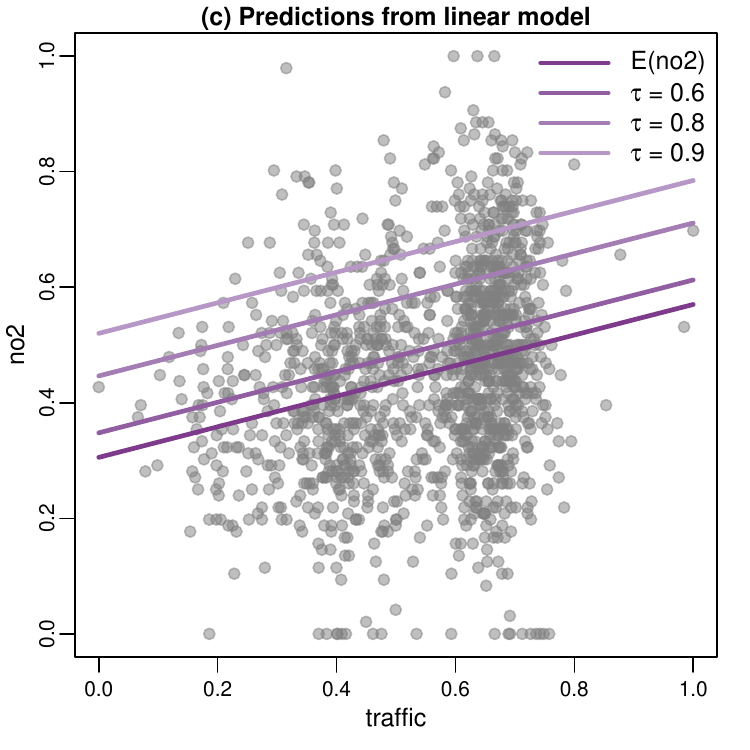}
\centering\includegraphics[width=0.48\textwidth,angle=0]{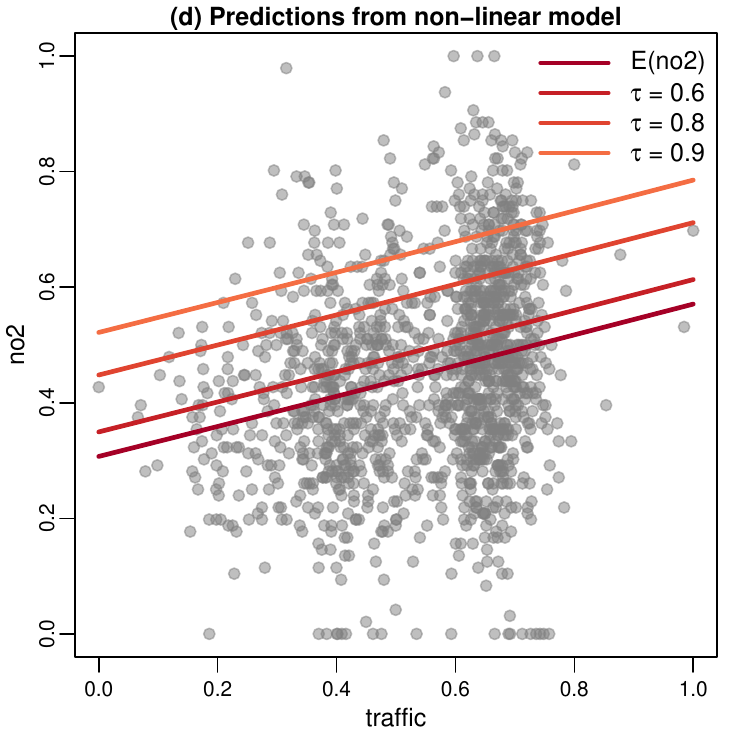}
\caption{Results from univariate preliminary analyses with $\mathit{traffic}$ as predictor. Panel (a) shows histograms of the response $\mathit{no2}$ according to the $\mathit{traffic}$-quartiles, i.e.~where $Q(\tau)$, $\tau\in\lbrace 0.25,0.50,0.75\rbrace$ is the corresponding quartile of the empirical $\mathit{traffic}$ distribution. Panel (b) shows normalized quantiles residuals from the univariate Gaussian regression model. Panels (c) and (d) show the predicted expectation $\dsE(\mathit{no2})$ as well as threshold quantiles $\tau\in\lbrace  0.6,0.8,0.9\rbrace$ obtained from a linear and nonlinear Gaussian model, respectively.}
\label{fig:prelim:no2:traffic}
\end{figure}

Figure \ref{fig:prelim:no2:o3} shows some preliminary univariate analyses with $\mathit{o3}$ as predictor. In particular, we show histograms of the response $\mathit{no2}$ according to the $\mathit{o3}$-quartiles, i.e.~where $Q(\tau)$, $\tau\in\lbrace 0.25,0.50,0.75\rbrace$ is the corresponding quartile of the empirical $\mathit{o3}$ distribution in panel (a). We also show normalized quantile residuals from the univariate Gaussian regression model (panel (b)). Then, we finally depict the predicted expectation $\dsE(\mathit{no2})$ as well as threshold quantiles $\tau\in\lbrace  0.6,0.8,0.9\rbrace$ obtained from a linear and nonlinear Gaussian model, respectively (panels (c) and (d)).
In the same line, Figure \ref{fig:prelim:no2:traffic} shows  initial results when using $\mathit{traffic}$ as a predictor {in a univariate mean regression model}. Looking at {these two} figures, we note that depending on the predictor values, NO$_2$ distributions differ (see panels (a)), {and looking at the tails, we note that NO$_2$ distributions are non-Gaussian} (see panels (b)). Furthermore, the quantiles of linear and nonlinear regressions are parallel, which is not appropriate because, conditional on traffic, the variance of NO$_2$ increases with increasing traffic. Finally, we observe that the predictor O$_3$ has a clear nonlinear effect on NO$_2$, while the one for traffic is rather linear.
We note that the apparent linearity we can observe between traffic and NO$_2$ in Figure \ref{fig:prelim:no2:traffic} is sort of spurious due to the increasing variability of the response, which makes the linearity not suitable. We refer to Section 5 where we further analyze this and indeed in Table~\ref{tab:inc}, we see that traffic also has a non-linear effect.

\section{Consistent effect selection for additive quantile regression}\label{sec:effselQR}
We first review STAQ models with a particular focus on semi-parametric predictors.  We then motivate the DR basis expansion over the commonly employed MMR. We show that the former which yields an orthogonal decomposition of each additive component into linear and nonlinear effect parts which cannot be achieved with the MMR. Finally, we establish consistency of the resulting effect estimators, providing the theoretical basis for quantile-specific linear and nonlinear effect estimation with the DR basis.

  We  consider the model formulation 
\begin{equation}\label{eq:mod}
y_i=\eta_{i,\tau}+\varepsilon_{i,\tau},
\end{equation}
where $\eta_{i,\tau}$ is a structured additive predictor~\citep{Woo2017} for a specific conditional quantile $\tau$ and $\varepsilon_{i,\tau}$ is an appropriate error term. Rather than assuming a zero mean for the errors as in classical mean regression, in quantile regression  we assume that the $\tau$-th quantile of the error term is zero, i.e., $F_{\varepsilon_{i,\tau}}(0) =\tau$, where $F_{\varepsilon_{i,\tau}}(\cdot)$ denotes the cumulative distribution function of the $i$th error term. This assumption implies that the predictor $\eta_{i,\tau}$ specifies the $\tau$-th quantile of $y_i$ and, as a result, the regression effects are directly interpretable at the conditional quantiles of the response distribution. Although our application only focuses on three specific quantiles, estimation for a dense set of quantiles would allow us to characterize the complete distribution of the responses in terms of covariates~\cite[see, e.g.][]{BonReiWan2010,SchEil2013,RodDorFan2019}.

\subsection{Semi-parametric predictors}\label{subsec:semipred}

We decompose the predictors $\eta_{i,\tau}$ in \eqref{eq:mod}  into $\eta_{i,\tau}=\eta_{i,\tau,\notsel} +\eta_{i,\tau,\mathrm{sel}}$, where the first sub-predictor  $\eta_{i,\tau,\notsel}$ contains effects not subject to selection, and the second sub-predictor $\eta_{i,\tau,\mathrm{sel}}$ contains effects subject to selection, respectively. We assume that $\eta_{i,\tau,\notsel}$ and $\eta_{i,\tau,\mathrm{sel}}$  are disjoint. Dividing the effects into two subsets enables certain covariate effects to be included by design, for instance, when prior knowledge supports their relevance or when they act as confounders that must be controlled for. In our application, treating \textit{year} as always included 
is justified because temporal changes in reporting practices and exposure 
cannot be ruled out. Hence, adjusting for \textit{year} prevents systematic 
temporal drift from being absorbed spuriously by other covariate effects.  Following the idea of STAR models, the predictor of observation $i$ is then 
\begin{equation*}
 \eta_{i,\tau} =  \beta_{\tau,0} + \sum_{k=1}^{K_\tau}f_{\tau,k,\notsel}(x_{i,k}) + \sum_{j=1}^{J_\tau}f_{\tau,j,\mathrm{sel}}(x_{i,j}),
\end{equation*}
where $\beta_{\tau,0}$ is the intercept term, while the effects $f_{\tau,k,\notsel}(x_{i,k})$ and $f_{\tau,j,\mathrm{sel}}(x_{i,j})$ represent flexible covariate effects.  In the following, we focus on the specification of effect selection  for $f_{\tau,j,\mathrm{sel}}(x_{i,j})$ of univariate continuous covariates $x_{i,j}$ since in our application no specific functional form of any of the nine covariates in \eqref{eq:predapp} should be excluded from selection a priori.  In contrast, as motivated above, we consider  $\mathit{year}$ as a fixed effect covariate (in addition to the intercept term), such that $\eta_{i,\tau,\notsel}=\beta_{\tau,0}+\sum_{k=1}^{K_\tau}\beta_{\tau,k}\mathit{year}_{i,k}$. If present, other information can be taken into account, such as spatial effects or random effects, see \citet{Woo2017} for possible further effect types.

\subsection{Effect decomposition}\label{subsect:effectdecomp}
Motivated by our application, we are interested in decomposing each effect subject to selection \(f_{\tau,j,\mathrm{sel}}\), \(j=1,\ldots,J_\tau\), into a linear and a
nonlinear component. To simplify notation, we suppress the subscript
``\(\mathrm{sel}\)'' throughout the remainder of this section and write
\(f_{\tau,j}\equiv f_{\tau,j,\mathrm{sel}}\). The decomposition is then given by
\[
 f_{\tau,j}(x_j)=f_{\tau,j,\unpen}(x_j)+f_{\tau,j,\pen}(x_j)=\widetilde{\beta}_{\tau,j} \widetilde{x}_j + \sum_{d=1}^{D_{\tau,j}}\beta_{\tau,j,d}B_{\tau,j,d}(x_j)
\]
 where $\widetilde{x}_j$  is an empirically standardized version of the original covariate vectors $x_j$, the functions $B_{\tau,j,d}(x_{i,j})$, $d\in\{1,\ldots,D_{\tau,j}\}$ denote suitable spline basis functions evaluated at $x_j$ capturing nonlinear covariate effects, and $\widetilde{\beta}_{\tau,j}$, $\beta_{\tau,j}=(\beta_{\tau,j,1},\ldots,\beta_{\tau,j,D_{\tau,j}})^\top$ are the unknown linear and nonlinear coefficients, respectively. Specifically, we obtain $\widetilde{x}_j$ via standardization with the empirical mean and standard deviation to approximate $\dsE(\widetilde{x}_j)=0$ and $\mbox{Var}(\widetilde{x}_j)=1$. 

 Recently, \citet{BacKle2024} showed that effect decomposition based on MMRs of penalized splines does not allow for consistent effect decomposition. In particular, for the spectral decomposition of the penalty matrix \citep[MMR;][]{FahKneLan2004}, or Eilers' transformation \citep{Eil1999}, the resulting spline basis functions are not empirically orthogonal to the linear functions. Consequently, the linear and nonlinear components cannot be 
separated consistently. To resolve  this issue, we follow \citet{BacKle2024} and expand the nonlinear effects in DR spline bases based on $\mathcal{L}^2(\dsP^{X_j})$-projections onto the linear functions $\mathcal U_j=\operatorname{span}\lbrace 1,X_j\rbrace,$ where  \(\mathcal{L}^2(\mathds P^{X_j})\) denotes the space of square-integrable functions with respect to the distribution of  $X_j$, $j\in\lbrace 1,\ldots, J_\tau\rbrace$. Specifically, for each continuous covariate \(j\), let
\[
  B_{\tau,j}^{\mathrm{raw}}(x_j)
  =
  \bigl(
    B_{\tau,j,1}^{\mathrm{raw}}(x_j),
    \ldots,
    B_{\tau,j,D_{\tau,j}+2}^{\mathrm{raw}}(x_j)
  \bigr)^\top
  \in\mathds R^{D_{\tau,j}+2}
\]
denote the cubic B-spline basis. For both the simulation study and the application, we choose $D_{\tau,j}=10$, which provides sufficient flexibility for the models considered. Furthermore, let  
\[
  B_{\tau,j}^{\mathrm{raw}}
  =
  \begin{pmatrix}
    B_{\tau,j}^{\mathrm{raw}}(x_{1,j})^\top\\
    \vdots\\
    B_{\tau,j}^{\mathrm{raw}}(x_{n,j})^\top
  \end{pmatrix}
  \in\mathds R^{n\times(D_{\tau,j}+2)}
\]
be the corresponding cubic B-spline design matrix. As it is common in the literature of penalized B-splines,   $B_{\tau,j}^{\mathrm{raw}}$ is matched with a second-order random-walk type penalty with roughness penalty $K_{\tau,j}^{\mathrm{raw}}=\Delta_j^\top\Delta_j$, where $\Delta_j\in \mathds{R}^{D_{\tau,j} \times (D_{\tau,j}+2)}$ with entries $\Delta_j[ik] = 
1  \text{ if } k = i,
-2 \text{ if } k = i+1, 
1 \text{ if } k = i+2, 
0  \text{ otherwise} \text{ for } i = 1, \dots, D_{\tau,j},\; k = 1, \dots, D_{\tau,j}+2$. 
The corresponding  DR basis vector  denoted by
\[
  B_{\tau,j}(x_j)
  =
  \bigl(
    B_{\tau,j,1}(x_j),
    \ldots,
    B_{\tau,j,D_{\tau,j}}(x_j)
  \bigr)^\top
  \in\mathds R^{D_{\tau,j}},
\]
with corresponding DR design matrix 
\[
  B_{\tau,j}
  =
  \begin{pmatrix}
    B_{\tau,j}(x_{1,j})^\top\\
    \vdots\\
    B_{\tau,j}(x_{n,j})^\top
  \end{pmatrix}
  \in\mathds R^{n\times D_{\tau,j}},
\]
 is obtained by applying Algorithm 1 of \citet{BacKle2024}. This algorithm performs a change of basis based on effect-specific transition matrices obtained from solving a generalized eigenvalue problem of the cross product of cubic B-splines and the penalty matrix. 

 The resulting design matrices $B_{\tau,j}$ 
have the following favorable properties which are not satisfied for the MMR of \citet{FahKneLan2004}.
\begin{enumerate}
\item Following Definition~1 of \citet{BacKle2024}, the linear component
      \(
      f_{\tau,j,\unpen}(x_j)
      \)
      is defined as the $\mathcal{L}^2(\mathds{P}^{X_j})$-projection of
      $f_{\tau,j}(x_j)$ onto $\mathcal U_j$. 
      The residual part,
      \(
      f_{\tau,j,\pen}(x_j)
      = f_{\tau,j}(x_j) - f_{\tau,j,\unpen}(x_j),
      \)
      represents the nonlinear component.
      \item The DR basis functions are empirically orthogonal to the linear functions, which allows consistent decomposition of linear and nonlinear components in STAQ models, as shown theoretically below.
 \item The DR basis functions are mutually orthogonal so that 
      $B_{\tau,j}^\top B_{\tau,j}$ is diagonal, and the associated roughness 
      penalties $K_{\tau,j}$ reduce to identity matrices. 
      This substantially improves computational efficiency compared to the MMR decomposition, as we show empirically in Section \ref{subsec:simulsum}. 
\end{enumerate}

\subsection{Consistent effect decomposition}\label{sec:theo}

In the following, we show that for any $\tau\in(0,1)$ we can consistently estimate the components  $ f_{\tau,j,\unpen}(x_j)$, $ f_{\tau,j,\pen}(x_j)$ in \eqref{eq:mod}. To do so, in a first step, we establish consistency of the  overall effects $f_{\tau,j}$ in our STAQ model in Corollary \ref{cor:overallQR}. The proof of the corollary verifies that our DR basis satisfies the  assumptions  of \citet{HorLee2005} who establish consistency  in nonparametric additive quantile 
regression models. In a second step, we combine Corollary \ref{cor:overallQR} with Theorem 1 of \citet{BacKle2024} to achieve consistent effect decomposition in our STAQ model with DR basis in Theorem \ref{thm:splitQR}.
 To this end, we first introduce some additional notation and make the following assumptions. For notational convenience, we shorten \cite{HorLee2005} to HL05 and \cite{BacKle2024} to BK24, respectively. 

\paragraph*{Additional notation and assumptions}
For each continuous covariate \(j\), we
define the univariate cubic
spline space as
\[
  \mathcal S_{\tau,j}
  =
  \left\{
    x_j\mapsto B_{\tau,j}^{\mathrm{raw}}(x_j)^\top a:
    a\in\mathds R^{D_{\tau,j}+2}
  \right\}.
\]
For functions \(h_1,h_2\in L^2(\mathds P^{X_j})\), write
\begin{align*}
  \langle h_1,h_2\rangle_j
  &=
  \dsE\{h_1(X_j)h_2(X_j)\}, 
  &\|h\|_j^2&=\langle h,h\rangle_j,\\
  \langle h_1,h_2\rangle_{n,j}
  &=
  \frac1n\sum_{i=1}^n h_1(x_{i,j})h_2(x_{i,j}), 
  &\|h\|_{n,j}^2&=\langle h,h\rangle_{n,j}.
\end{align*}
for the theoretical and empirical semi-inner product on $ L^2(\mathds P^{X_j})$. 
Let \(H_j\) be the \(L^2(\mathds P^{X_j})\)-projector onto \(\mathcal U_j\), and let
\(H_{n,j}\) be the corresponding empirical projector:
\[
  H_jh
  =
  \arg\min_{u\in\mathcal U_j}\|h-u\|_j^2,
  \qquad
  H_{n,j}h
  =
  \arg\min_{u\in\mathcal U_j}\|h-u\|_{n,j}^2.
\]
Set
\(
  R_j=I-H_j,
  \;
  R_{n,j}=I-H_{n,j}, 
\)
such that
\(
  f_{\tau,j,\unpen}=H_j f_{\tau,j},\;
  f_{\tau,j,\pen}=R_j f_{\tau,j}.
\)
We make the following assumptions.
\begin{assumption}[Regularity conditions]
\label{assumption:A}\

\begin{enumerate}[label=(A\arabic*), ref=(A\arabic*)]
  \item\label{assumption:itemA1}
 Let \((Y_i,X_i)_{i=1}^n\) be i.i.d.~observations, where $X=(X_{1},\ldots,X_{J_\tau})^\top$. The true conditional \(\tau\)-quantile function is additive and admits the decomposition
  \[
    Q_{Y \mid X}(\tau \mid X)
    =
    \beta_{\tau,0}^* + 
    \sum_{j=1}^{J_\tau} f_{\tau,j}^*(X_{j}),
  \]
  where \(\beta_{\tau,0}^*\) is the true intercept and \(f_{\tau,j}^*\) denotes the true
component function associated with the \(j\)-th continuous covariate (which can also be zero if not truly active). Without loss of generality, the component functions are centered so that
\(\mathbb{E}\{f_{\tau,j}^*(X_j)\}=0\) for all \(j=1,\ldots,J_\tau\),
ensuring identifiability of the additive decomposition.

  \item\label{assumption:itemA2}
  Each continuous covariate \(X_j\) has compact support, normalized to
  \([0,1]\), and density bounded away from zero and infinity.

  \item\label{assumption:itemA3}
  The residual
  \(
    U_\tau
    =
    Y- Q_{Y \mid X}(\tau \mid X)
  \)
  satisfies \(P(U_\tau\le0\mid X)=\tau\). Its conditional density exists in a
  neighborhood of zero, is Lipschitz in its first argument, and is bounded away from zero
  and infinity uniformly in \(X\).

  \item\label{assumption:itemA4}
  Each \(f_{\tau,j}^*\in C^r([0,1])\) for some \(r>2\), and the spline degree is at least
  \(r-1\). In particular, our default cubic choice $m=3$ accommodates $r\in(2,4]$. The number of nonlinear DR basis functions satisfies
  \[
    D_{\tau,j}\to\infty,
    \qquad
    D_{\tau,j}=o(\sqrt n),
    \qquad
    \frac{D_{\tau,j}^4(\log n)^2}{n}\to0,
    \qquad
    \frac{D_{\tau,j}^{1+2r}}{n}=O(1).
  \]

  \item\label{assumption:itemA5}
  Let
    \[
      G_{\tau,j}
      =
      \mathbb{E}\!\left[
      B_{\tau,j}(X_j)
      B_{\tau,j}(X_j)^\top
      \right]
    \]
    denote the population Gram matrix associated with the DR basis
    \(B_{\tau,j}(\cdot)\). Assume that the eigenvalues of \(G_{\tau,j}\) are bounded
    away from zero and infinity uniformly in \(j\). In addition, assume that the corresponding
    empirical Gram matrix
    \[
      \widehat G_{\tau,j}
      =
      \frac1n
      \sum_{i=1}^n
      B_{\tau,j}(x_{i,j})
      B_{\tau,j}(x_{i,j})^\top
    \]
    is uniformly nonsingular with probability tending to one.

  \item\label{assumption:itemA6}
  For each fixed \(\tau\in(0,1)\) and \(j=1,\ldots,J_\tau\), the DR basis \(  B_{\tau,j,1},\ldots,B_{\tau,j,D_{\tau,j}} \) satisfies the following conditions on an
  event whose probability tends to one:
  \begin{enumerate}[label=(\alph*)]
    \item
    Each \(B_{\tau,j,d}\), \(d=1,\ldots,D_{\tau,j}\), is continuous.

    \item
    The DR basis functions are empirically orthogonal to the linear functions:
    \[
      \frac1n\sum_{i=1}^n B_{\tau,j,d}(x_{ij})=0,
      \qquad
      \frac1n\sum_{i=1}^n \widetilde x_{ij} B_{\tau,j,d}(x_{ij})=0,
      \qquad
      d=1,\ldots,D_{\tau,j}.
    \]

    \item
    The DR basis functions are mutually empirically orthogonal:
  \[
    \frac1n\sum_{i=1}^n
    B_{\tau,j,d}(x_{ij})B_{\tau,j,d'}(x_{ij})
    =
    0,
    \qquad
    d\neq d',\quad
    d,d'=1,\ldots,D_{\tau,j}.
  \]

    \item
    As \(D_{\tau,j}\to\infty\),
    \[
      \sup_{x_j\in[0,1]}\|B_{\tau,j}(x_j)\|
      =
      O_p(D_{\tau,j}^{1/2}).
    \]

    \item
    There exist 
    \(\widetilde\beta_{\tau,j}\in\mathds R\), and
    \(\beta_{\tau,j}\in\mathds R^{D_{\tau,j}}\) such that
    \[
      \sup_{x_j\in[0,1]}
      \left|
        f_{\tau,j}^*(x_j)
        -
        \widetilde\beta_{\tau,j} \widetilde x_j
        -
        B_{\tau,j}(x_j)^\top\beta_{\tau,j}
      \right|
      =
      O(D_{\tau,j}^{-r}).
    \]
  \end{enumerate}
\end{enumerate}
\end{assumption}

Conditions \ref{assumption:itemA1}--\ref{assumption:itemA3} are common quantile regression regularity
conditions. Condition~\ref{assumption:itemA4} ensures that the true component functions are
smooth enough and that the spline dimension grows at an admissible rate. Condition
\ref{assumption:itemA5} ensures that the resulting DR spline design is
well-conditioned. The assumptions above impose the HL05-type regularity conditions that are not specific to the choice of basis. Since our estimator is based on the DR basis, it remains to verify Assumption~\ref{assumption:itemA6}. This is shown in the Supplementary Material A.2. 
We therefore obtain the following Corollary.

\begin{corollary}[Overall consistency of the DR additive quantile effect]
\label{cor:overallQR}
Suppose Assumption~\ref{assumption:A} holds. For $\tau\in(0,1)$, let
\[
  \widehat f_{\tau,j}(x_j)
  =
  \widehat f_{\tau,j,\unpen}(x_j)
  +
  \widehat f_{\tau,j,\pen}(x_j)
  =
  \widehat{\widetilde\beta}_{\tau,j}\widetilde x_j
  +
  B_{\tau,j}(x_j)^\top\widehat\beta_{\tau,j}
\]
be the DR spline estimator, where we assume that \( \widehat f_{\tau,j} =  \left(\widehat f_{\tau,j,n}\right)_{n \in \mathds{N}} \) is the sequence of random spline functions that are centered empirically at covariate $j$. Then, for every
\(j=1,\ldots,J_\tau\),
\[
  \bigl\|
    \widehat f_{\tau,j}
    -
    f_{\tau,j}^*
  \bigr\|_{j}
  \xrightarrow{P}
  0.
\]
\end{corollary}
The proof is given in the Supplementary Material A.3.

\begin{theorem}[Consistent linear and nonlinear DR effect decomposition]
\label{thm:splitQR}
Suppose Assumption~\ref{assumption:A} holds and let $\tau\in(0,1)$. For each  
continuous covariate \(j\in\{1,\ldots,J_\tau\}\), define
\[
  f_{\tau,j,\unpen}^*
  =
  H_j f_{\tau,j}^*,
  \qquad
  f_{\tau,j,\pen}^*
  =
  R_j f_{\tau,j}^*.
\]
Let
\[
  \widehat f_{\tau,j,\unpen}(x_j)
  =
  \widehat{\widetilde\beta}_{\tau,j}\widetilde x_j,
  \qquad
  \widehat f_{\tau,j,\pen}(x_j)
  =
  B_{\tau,j}(x_j)^\top\widehat\beta_{\tau,j}
\]
be the fitted DR linear and nonlinear components. Then, for every
\(j=1,\ldots,J_\tau\),
\[
  \bigl\|
    \widehat f_{\tau,j,\unpen}
    -
    f_{\tau,j,\unpen}^*
  \bigr\|_{j}
  \xrightarrow{P}
  0,
  \qquad
  \bigl\|
    \widehat f_{\tau,j,\pen}
    -
    f_{\tau,j,\pen}^*
  \bigr\|_{j}
  \xrightarrow{P}
  0.
\]
\end{theorem}
The proof is given in the Supplementary Material A.4.

\section{Data-driven effect selection through Bayesian inference} \label{sec:bayes}
The preceding section motivates the
DR basis from the perspective of consistent effect decomposition. In this section, we
combine this representation with the asymmetric Laplace working likelihood and NBPSS
priors to obtain a data-driven Bayesian framework for quantile-specific effect selection,
estimation, prediction, and uncertainty quantification. We denote our method by \modname. 

\subsection{Bayesian  quantile regression}

To facilitate Bayesian estimation in quantile regression, we follow a common strategy in the literature and  assume an asymmetric Laplace distribution (ALD) $y_i\sim\mbox{ALD}(\eta_{i,\tau},\delta^{2},\tau)$ with location predictor $\eta_{i,\tau}$, scale parameter $\delta^{2}$, asymmetry parameter $\tau$, and density
\[
p(y_i|\eta_{i,\tau},\delta^{2},\tau)=\tau(1-\tau)\delta^{2}\exp(-\delta^{2}\rho_\tau(y_i-\eta_{i,\tau}))
\]
as working likelihood. This assumption is reasonable because it can be shown that minimizing the loss function \eqref{eq:awad} is equivalent to maximizing the likelihood function induced by the ALD.
However, estimation with the ALD directly is not straightforward due to the non-differentiability of the check function $\rho_\tau(\cdot)$ at zero. To address this,  several authors considered rewriting the ALD as a scale mixture of two Gaussian distributions~\citep{ReeYu2009,KozKob2011,YueRue2011,LumGel2012}.  Specifically, \citet{Tsi2003} shows that assuming $Y\sim\mbox{ALD}(\eta,\delta^{2},\tau)$, $Y$ has the representation
\[\label{eq:M1a}
Y=\eta+\xi W + \sigma Z\sqrt{\delta^{-2} W},\quad W\sim\ExpD(\delta^{2}),\;Z\sim\ND(0,1), 
\]
where $\xi=\tfrac{1-2\tau}{\tau(1-\tau)}$, $\sigma^2=\tfrac{2}{\tau(1-\tau)}$ are two scalars depending on the quantile level $\tau$.  The random variables $W$ and $Z$ are independently distributed according to exponential ($\ExpD(\delta^{2})$ with rate $\delta^2$) and standard Gaussian ($\ND(0,1)$) distributions, respectively. We assume a conjugate gamma prior distribution $\delta^2\sim\GaD(a_\delta,b_\delta)$ with hyperparameters fixed at $a_\delta=b_\delta=0.001$. This representation as a mixture facilitates an easy way to construct Gibbs sampling for MCMC inference. In addition, \citet{YueRue2011} show empirically that the ALD can capture various data distributions. Finally, even though the ALD induces misspecification,  our simulations in the Supplementary Material C provide  empirical evidence that \modnamesp still allows for consistent effect selection and reasonable uncertainty quantification via the respective posterior samples.

\paragraph*{Alternative working likelihoods}
To perform Bayesian quantile regression, alternative working likelihoods have been considered. For instance, \citet{KotKrn2009} construct a generic class of
semi-parametric and non-parametric distributions for the likelihood using Dirichlet process mixture models, while \citet{ReiBonWan2010} consider a flexible infinite mixture of Gaussian combined with a stick-breaking construction for the priors. \citet{YangHe2012} propose  the Bayesian empirical likelihood approach to quantile regression.

\subsection{Hierarchical spike and slab prior for effect selection}\label{subsec:NBPSS}

We combine the DR basis with a Bayesian group-selection prior to automatically
identify which covariates have a linear effect, a nonlinear effect, both, or no
effect at all. To make this explicit, we rewrite the additive predictor as
\[
\eta_{i,\tau} 
= \eta_{i,\tau,\mathrm{in}} 
  + \sum_{j=1}^{2J_\tau} \nu_{\tau,j}(x_{i,j})\,\beta_{\tau,j},
\]
where each design vector $\nu_{\tau,j}(x_{i,j})$ corresponds either to a linear effect
(based on $\widetilde{x}_{i,j}$) or to a nonlinear effect (based on the $i$-th row of $B_{\tau,j}$). In this representation, effect selection becomes a group-sparsity problem:
some coefficient groups $\beta_{\tau,j}$ may shrink to zero, implying the 
absence of the corresponding effect. We place an NBPSS prior \citep{KleCarKneLanWag2021} on 
each $\beta_{\tau,j}$:
\begin{equation}
\label{eq:M5}	
 \begin{aligned}
 \beta_{\tau,j}\mid \zeta_{\tau,j}^2 & \sim\mathcal{N}_{D_{\tau,j}}\left(0,\zeta_{\tau,j}^2 I_{D_{\tau,j}}\right)\\
 \zeta_{\tau,j}^2\mid \gamma_{\tau,j},\psi_{\tau,j}^2 & \sim\GaD\left(\frac{1}{2},\frac{1}{2r_{\tau,j}(\gamma_{\tau,j}) \psi_{\tau,j}^2}\right) \\
 \gamma_{\tau,j}\mid \omega_{\tau,j}   & \sim \BerD(\omega_{\tau,j})\\
 \psi_{\tau,j}^2    & \sim \IGD(a_{\tau,j},b_{\tau,j})\\
 \omega_{\tau,j}    & \sim \BetaD(a_{0,\tau,j},b_{0,\tau,j})\\
 r_{\tau,j}\equiv r(\gamma_{\tau,j}) &=\begin{cases} r_{\tau,j}>0 \mbox{ small } & \gamma_{\tau,j}=0\\
                          1  & \gamma_{\tau,j}=1,
            \end{cases}
 \end{aligned}
\end{equation}
where $\mathcal{N}_{D_{\tau,j}}(\mu,\Sigma)$ denotes a $D_{\tau,j}$-dimensional
Gaussian distribution with mean vector $\mu$ and covariance matrix $\Sigma$;
$\GaD(a,b)$ is a gamma distribution with shape and scale parameters $(a,b)$;
$\BerD(p)$ is a Bernoulli distribution with success probability $p$;
$\IGD(a,b)$ is an inverse gamma distribution with parameters $(a,b)$; and
$\BetaD(a,b)$ is a beta distribution with shape parameters $(a,b)$.

The hierarchical structure of the NBPSS prior induces strong shrinkage toward
zero for groups with $\gamma_{\tau,j}=0$ through the small parameter 
$r_{\tau,j}(\gamma_{\tau,j})$ in the spike, while allowing substantial flexibility for
groups with $\gamma_{\tau,j}=1$, for which the variance component 
$\zeta_{\tau,j}^2$ is governed primarily by the slab distribution. 
In the DR basis representation, this separation becomes particularly effective:
since the linear and nonlinear components of each covariate are empirically 
orthogonal, the prior can shrink the nonlinear component without affecting the 
linear part and vice versa. 
This results in a clean and interpretable decomposition, enabling the model to 
distinguish linear effects, nonlinear effects, both simultaneously, or the 
absence of any effect. 
The hyperpriors on $\omega_{\tau,j}$ and $\psi_{\tau,j}^2$ provide additional 
adaptivity, allowing the sparsity level and slab variance to be learned from 
the data, which stabilizes estimation across quantiles and avoids the need for 
quantile-specific tuning.

For the intercept terms and linear effects not subject to selection (i.e., the $\mathit{year}$ effects), we use independent $\mathcal{N}(0,10^{10})$ distributions as priors.

\subsection{Prior elicitation}\label{subsec:PriEli}
In \eqref{eq:M5}, the hyperparameters $a_{\tau,j}$, $b_{\tau,j}$,
$a_{0,\tau,j}$, $b_{0,\tau,j}$, and $r_{\tau,j}$ determine the behavior of the
NBPSS prior. Following \citet{KleCarKneLanWag2021}, we set $a_{\tau,j}=1/2$ and
$b_{\tau,j}=5$. Since no specific prior knowledge about inclusion probabilities
is available in our application, we use $a_{0,\tau,j}=b_{0,\tau,j}=1$, which
corresponds to a uniform prior on $(0,1)$. In general, prior beliefs about the
inclusion probability can be encoded by choosing
$\mathds{P}(\gamma_{\tau,j}=1)=a_{0,\tau,j}/(a_{0,\tau,j}+b_{0,\tau,j})$.

\paragraph*{Hyperparameters $b_{\tau,j}$ and $r_{\tau,j}$}
To elicit $b_{\tau,j}$ and $r_{\tau,j}$, we adopt a streamlined version of the
approach in \citet{KleCarKneLanWag2021}, which itself builds on ideas from
\citet{SimRueMarRieSor2017} and \citet{KleKne2016}. The key idea is to phrase
prior information in terms of the supremum norm $\|\nu_{\tau,j}\beta_{\tau,j}\|_\infty$ of a specific group (and thus functional effect) in the predictor. Conditional on $\gamma_{\tau,j}=1$ (inclusion), the hyperparameter
$r_{\tau,j}$ drops out of the prior for $\nu_{\tau,j}\beta_{\tau,j}$, so $b_{\tau,j}$ can be
determined from the marginal probability
\begin{equation}\label{eq:cond:hyper2}
 \mathds{P}\!\left(\left.\|\nu_{\tau,j}\beta_{\tau,j}\|_\infty\leq c_{\tau,j}\,\right|\,\gamma_{\tau,j}=1\right)
 = \alpha_{\tau,j}.
\end{equation}
Here, $c_{\tau,j}$ and $\alpha_{\tau,j}$ should be chosen small to reflect the
belief that if the effect is truly active, the event
$\|\nu_{\tau,j}\beta_{\tau,j}\|_\infty\le c_{\tau,j}$ is unlikely. Analogously, $r_{\tau,j}$ is
chosen based on
\begin{equation}\label{eq:cond:hyper1}
 \mathds{P}\!\left(\left.\|\nu_{\tau,j}\beta_{\tau,j}\|_\infty\leq c_{\tau,j}\,\right|\,\gamma_{\tau,j}=0\right)
 = 1-\alpha_{\tau,j},
\end{equation}
where the probability is high because the excluded effects should remain small.

 To evaluate the
probabilities in \eqref{eq:cond:hyper2}--\eqref{eq:cond:hyper1}, we simulate
from the marginal prior of $\|\nu_{\tau,j}\beta_{\tau,j}\|_\infty$. While
\citet{KleCarKneLanWag2021} solve these conditions numerically via
optimization, we note that this can be  substantially more efficient using
a scaling property of the scaled beta prime distribution:
\citet{PerPerRam2017} show that the marginal priors of the spike and slab
components $p(\zeta_{\tau,j}^2 \mid \gamma_{\tau,j})$ are scaled beta prime
distributions with shape parameters $(1/2, a_{\tau,j})$ and scale parameter
$2r_{\tau,j}b_{\tau,j}$. Thus,
$\widecheck\zeta_{\tau,j}^2 = \zeta_{\tau,j}^2 / (2r_{\tau,j}b_{\tau,j})$ satisfies
$\widecheck\zeta_{\tau,j}^2 \mid \gamma_{\tau,j} \sim \mathrm{BP}(1/2, a_{\tau,j})$.
Writing $\nu_{\tau,j}\beta_{\tau,j} = \sqrt{2r_{\tau,j}b_{\tau,j}}\,
\widecheck\zeta_{\tau,j}\,\nu_{\tau,j}\widecheck\beta_{\tau,j}$ for $\widecheck\beta_{\tau,j}=\beta_{\tau,j}/\zeta_{\tau,j}$ shows that
\eqref{eq:cond:hyper2} is equivalent to
\[
\mathds{P}\!\left(
 \left.
 \|\widecheck\zeta_{\tau,j} \nu_{\tau,j}\widecheck\beta_{\tau,j}\|_\infty
 \le c_{\tau,j}/\sqrt{2b_{\tau,j}}
 \,\right|\,\gamma_{\tau,j}=1
\right)=\alpha_{\tau,j},
\]
and similarly for \eqref{eq:cond:hyper1} with $r_{\tau,j}$ in place of~$1$. Algorithm 1 in the Supplementary Material B.1 gives a pseudo algorithm for practical computation. 

The choices of $c_{\tau,j}$ and $\alpha_{\tau,j}$ control the relative balance
between true  and false positives. Smaller values lead to sparser
models. Based on our simulation studies, we use the default
$c_{\tau,j}=0.1$ and $\alpha_{\tau,j}=0.01$. However, we highlight that if certain prior knowledge is available, it can be useful to make these values not only effect-specific (index $j$) but also quantile-specific (index $\tau$).

\subsection{Posterior inference}\label{subsec:post}

Posterior sampling for \modnamesp can be realized in Gibbs updates as follows. For $\tau\in(0,1)$, let
\begin{equation*}\begin{aligned}
    \theta_{\tau}&=\lbrace \beta_{\tau,1},\ldots,\beta_{\tau,2 J_{\tau}},\zeta_{\tau,1}^2,\ldots,\zeta_{\tau,2 J_{\tau}}^2,\gamma_{\tau,1},\ldots,\gamma_{\tau,2 J_{\tau}},\psi_{\tau,1},\ldots,\psi_{\tau,2 J_{\tau}},\omega_{\tau,1},\ldots,\omega_{\tau,2 J_{\tau}},\\
    &\qquad w_1,\ldots,w_n,\delta^2, \beta_{\tau,0},\beta_{\tau,1,\notsel},\ldots,\beta_{\tau,K_{\tau},\notsel}\rbrace
    \end{aligned}\end{equation*}
    be the set of all model parameters, where w.l.o.g.~we assume that only the intercept and the year-specific effects are not subject to selection to match our actual predictor in \eqref{eq:predapp} for our application. Then, denoting by $p(\cdot\mid\theta_\tau\backslash\cdot,y)$ the conditional posterior distribution of $(\cdot)$ given all other parameters, the MCMC sampler can be summarized as follows.

\paragraph*{MCMC sampler for \modname}
 For a given quantile $\tau\in(0,1)$ and iteration $s=1,\ldots,S$ of the MCMC sampler:\\
 
\noindent \underline{Step~1.} For $j=1,\ldots,2J_\tau$ generate from $p(\beta_{\tau,j}\,|\,\theta_{\tau}\backslash\beta_{\tau,j},y)$.\\
\noindent \underline{Step~2.} For $j=1,\ldots,2J_\tau$ generate from $p(\zeta_{\tau,j}^2\,|\,\theta_{\tau}\backslash\zeta_{\tau,j},y)$.\\
\noindent \underline{Step~3.} For $j=1,\ldots,2J_\tau$ generate from $p(\gamma_{\tau,j}\,|\,\theta_{\tau}\backslash\gamma_{\tau,j},y)$.\\
\noindent \underline{Step~4.} For $j=1,\ldots,2J_\tau$ generate from $p(\psi_{\tau,j}^2\,|\,\theta_{\tau}\backslash\psi_{\tau,j},y)$.\\
\noindent \underline{Step~5.} For $j=1,\ldots,2J_\tau$ generate from $p(\omega_{\tau,j}\,|\,\theta_{\tau}\backslash\omega_{\tau,j},y)$.\\
\noindent \underline{Step~6.} For $i=1,\ldots,n$ generate from $p(w_i\,|\,\theta_{\tau}\backslash w_i,y)$. \\
\noindent \underline{Step~7.} Generate from $p(\delta^2\,|\,\theta_{\tau}\backslash\delta^2,y)$.\\

At Step 1, the full conditional distributions for $\beta_{\tau,j}$ are Gaussian distributions $\ND(\mu_{\tau,j},\Sigma_{\tau,j})$ due to the conjugate model hierarchy implied by the location-scale mixture representation and the Gaussian priors for $\beta_{\tau,j}$. The mean $\mu_{\tau,j}$ and covariance $\Sigma_{\tau,j}$ are:
\begin{equation*}\begin{aligned}
 \mu_{\tau,j}&=\Sigma_{\tau,j}\left(\tfrac{\delta^2}{\sigma^2}\nu_{\tau,j}^\top D_w^{-1}(y-\xi w-(\eta_{\tau}-\nu_{\tau,j}\beta_{\tau,j}))\right)
\\
\Sigma_{\tau,j}&=\left(\zeta_{\tau,j}^2 I_{D_\tau,j}+\tfrac{\delta^2}{\sigma^2}\nu_{\tau,j}^\top D_w^{-1}\nu_{\tau,j}\right)^{-1},
\end{aligned}\end{equation*}
where $w=(w_1,\ldots,w_n)^\top$ and $\mD_w=\mbox{diag}(w_1,\ldots,w_n)$.

At Step 2, we note that $p(\zeta_{\tau,j}^2|\beta_{\tau,j},\gamma_{\tau,j},\psi_{\tau,j}^2)$ is a generalized inverse Gaussian distribution $\mathcal{GIG}(p,q,c)$, with $p=-0.5 D_{\tau,j}+0.5$, $q=1/(r_{\tau,j}\psi_{\tau,j}^2)$, $c=\beta_{\tau,j}^\top \beta_{\tau,j}$. 

For Steps 3--5 it is easy to show that 
\[
\left. p(\gamma_{\tau,j}=1\,\right|\,\theta_\tau\backslash\gamma_{\tau,j},y) = \left(1+\frac{\varphi(\zeta_{\tau,j};0; r_{\tau,j}\psi_{\tau,j}^2) (1-\omega_{\tau,j})}{\varphi(\zeta_{\tau,j};0;\psi_{\tau,j}^2)\omega_{\tau,j}} \right)^{-1},
\]
 where $\varphi(\cdot; \mu, \sigma^2)$ denotes the density of a univariate Gaussian distribution with mean $\mu$ and variance $\sigma^2$; and
\begin{equation*}\begin{aligned}
 \left.\psi_{\tau,j}^2\,\right|\, \theta_\tau\backslash\psi_{\tau,j}^2,y &\sim \IGD\left(a_{\tau,j}+0.5,b_{\tau,j}+\frac{\zeta_{\tau,j}^2}{2r_{\tau,j}}\right)
\\
 \left.\omega_{\tau,j}\,\right|\,\theta_\tau\backslash\omega_{\tau,j},y &\sim \BetaD(a_{0,\tau,j}+\gamma_{\tau,j},b_{0,\tau,j}+1-\gamma_{\tau,j}),
\end{aligned}\end{equation*}

To generate $w_i$, $i=1,\ldots,n$ at Step 6, we  note that the full conditional distribution for the weights are  generalized inverse Gaussian:
\[
\left.w_i\,\right|\,\theta_\tau\backslash w_i,y\sim\mathcal{GIG}\left(\frac{1}{2},\frac{\delta^2(y_i-\eta_{i,\tau})^2}{\sigma^2},\frac{\delta^2(\xi^2+2\sigma^2)}{\sigma^2}\right).
\]

At Step 7, the full conditional distribution for $\delta^2$  is a  gamma distribution:
\[
\left.\delta^2\,\right|\,\theta_\tau\backslash \delta^2,y\sim\GaD\left(a_\delta+\frac{3n}{2},b_\delta+\frac{1}{2\sigma^2}\sum_{i=1}^n\frac{(y_i-\eta_{i,\tau}-\xi w_i)^2}{w_i}+\sum_{i=1}^n w_i\right).
\]

\subsection{Variable selection, prediction, and uncertainty quantification}\label{sec:subpred}

For $\tau\in(0,1)$, let $\left\lbrace \theta_\tau^{[1]},\ldots,\theta_\tau^{[S]}\right\rbrace$ denote an MCMC sample from the joint posterior of $\theta_\tau$, where for $s=1,\ldots,S$,
\begin{equation*}\begin{aligned}
    \theta_{\tau}^{[s]}&=\lbrace \beta_{\tau,1}^{[s]},\ldots,\beta_{\tau,2 J_{\tau}}^{[s]},(\zeta_{\tau,1}^2)^{[s]},\ldots,(\zeta_{\tau,2 J_{\tau}}^2)^{[s]},\gamma_{\tau,1}^{[s]},\ldots,\gamma_{\tau,2 J_{\tau}}^{[s]},\psi_{\tau,1}^{[s]},\ldots,\psi_{\tau,2 J_{\tau}}^{[s]},\omega_{\tau,1}^{[s]},\ldots,\omega_{\tau,2 J_{\tau}}^{[s]},\\
    &\qquad w_1^{[s]},\ldots,w_n^{[s]},(\delta^2)^{[s]},\beta_{\tau,0}^{[s]},\ldots,\beta_{\tau,K_\tau,\notsel}^{[s]}\rbrace.
    \end{aligned}\end{equation*}
    Furthermore, let $j\in\lbrace 1,\ldots, 2 J_\tau\rbrace$.

\paragraph*{Variable selection}
We perform variable selection based on the Rao-Blackwellized estimator
\begin{equation}\label{eq:pip}
    \widehat p_{\tau,j}=\frac{1}{S}\sum_{s=1}^S \dsP(\gamma_{\tau,j}^{[s]}=1\mid \theta_{\tau}^{[s]}\backslash\gamma_{\tau,j}^{[s]} ,y)
\end{equation}
 and say that a specific functional effect (linear or nonlinear) is included in the model if $\hat p_{\tau,j}\geq 0.5$.

\paragraph*{Point estimates}
We define point estimates for each functional  effect $f_{\tau,j}$ as 
\begin{equation}\begin{aligned}\label{eq:postmean}
 \widehat{f}_{\tau,j}(x_j)&=\frac{1}{S}\sum_{s=1}^S \nu_{\tau,j}(x_j)\beta_{\tau,j}^{[s]}  \\
 &=\begin{cases}\frac{1}{S}\sum_{s=1}^S\beta_{\tau,j}^{[s]} \widetilde{x}_j &\mbox{ if } j\leq J_\tau\\
 \frac{1}{S}\sum_{s=1}^S\left(\sum_{d=1}^{D_{\tau,j}}\beta_{\tau,J_\tau+j,d}^{[s]} B_{\tau,j,d}(x_j)\right) &\mbox{ if } J_\tau<j\leq 2 J_\tau.
 \end{cases}
\end{aligned}\end{equation}
Furthermore, we define the point estimates for the conditional quantile  $Q_{Y \mid X}(\tau\mid X)=\eta_{\tau}$ as
\begin{equation*}\begin{aligned}
 \widehat{\eta}_{\tau}(x)=\widehat\beta_{\tau, 0}+\sum_{j=1}^{2J_\tau}\widehat{f}_{\tau,j}(x_j),
\end{aligned}\end{equation*}
where $x=(x_1,\ldots,x_{2J_\tau})^\top$. 
\paragraph*{Uncertainty quantification} Uncertainty for each effect is quantified pointwise using empirical $(1-\alpha)$ credible intervals derived from the posterior samples, with $\alpha=0.05$: 
\begin{equation}\begin{aligned}\label{eq:ci}
    \widehat{\mathrm{CI}}_{1-\alpha}(f_{\tau,j}(x_j))&=
\bigl[
\hat{\ell}_{\tau,j}(x_j),\, \hat{u}_{\tau,j}(x_j)
\bigr] \\
\hat{\ell}_{\tau,j}(x_j) &= Q_{\alpha/2}(\lbrace \nu_{\tau,j}(x_j)\beta_{\tau,j}^{[1]} ,\ldots,\nu_{\tau,j}\beta_{\tau,j}^{[S]} \rbrace) \\
\hat{u}_{\tau,j}(x_j) &= Q_{1-\alpha/2}(\lbrace \nu_{\tau,j}(x_j)\beta_{\tau,j}^{[1]} ,\ldots,\nu_{\tau,j}(x_j)\beta_{\tau,j}^{[S]} \rbrace).
\end{aligned}\end{equation}

In Section \ref{sec:results} and Supplementary Material C, estimated functions are evaluated pointwise on a fine grid  $x_j =0,0.01,\ldots,1$.

\subsection{Empirical evaluation}\label{subsec:simulsum}
In addition to the robustness checks of \modnamesp with regard to the ALD assumption, we  conduct a benchmark study to evaluate \modname. We consider various data generating processes, quantile levels and covariate designs. We compare \modnamesp to the competing MMR approach in terms of  estimation accuracy, variable selection performance, and uncertainty quantification. 
For both approaches, we also examine the sensitivity to the prior elicitation hyperparameters $(\alpha_{\tau,j}, c_{\tau,j})$. We document these in detail in the Supplementary Material C.

Taken together, the simulation results provide strong and consistent
empirical evidence that \modnamesp is generally favorable over  MMR. \modnamesp demonstrates accurate
variable selection, comparable or lower estimation error for functional effects,
sharper and better calibrated credible intervals, and lower interval
scores. 

Finally, \modnamesp is also computationally more efficient than MMR. To provide evidence for this, we record the runtime required to generate 4{,}000 posterior MCMC iterations after a burn-in of 1{,}000 iterations, without thinning, settings that we also consider in the simulation study and application. Table~\ref{tab:runtime_comparison} summarizes the runtimes for both methods for sample sizes $n = 100$, $1{,}000$, $10,000$, $50{,}000$, and $100{,}000$.   Across all  sample sizes considered, \modnamesp requires consistently less computational time than MMR, with a speed-up between approximately 5\% and 12\%. 
\begin{table}[htbp]
\centering\renewcommand\arraystretch{1.00}
\begin{tabular}{c|ccc}
  \hline\hline
 Sample Size ($n$) & \modnamesp (s) & MMR (s) & Speedup (\%) \\ 
  \hline
    100      & 5.454   & 6.193   & 11.93 \\
    1,000    & 42.475  & 44.529  & 4.61  \\
    10,000    & 418.81  & 448.84  & 6.7  \\
    50,000   & 2093.543 & 2196.930 & 4.71 \\
    100,000  & 4216.037 & 4451.960 & 5.30 \\
  \hline \hline
\end{tabular}
\caption{Observed runtimes (in seconds) for a single simulation replicate of \modnamesp and MMR across different sample sizes. The final column reports the relative speed improvement of \modnamesp compared to MMR. All experiments were conducted on a machine with an Apple M3 chip (14-inch MacBook Pro, 16 GB RAM) running macOS Tahoe Version 26.0.1.}
\label{tab:runtime_comparison}
\end{table}

\section{Analysis on Madrid's air pollution}\label{sec:results}

\paragraph*{Model specification}
We recall that Table \ref{tab:data} shows a description and summary statistics of  continuous covariates (before standardization to $[0,1]$) and from which we can extract the full predictor specification for a given quantile $\tau \in \{0.6, 0.8, 0.9\}$
\begin{equation*}\begin{aligned}
\eta_{\tau}&=\sum_{k=0}^4 \beta_{\tau,k} + f_{\tau,1}(\mathit{co}) + f_{\tau,2}(\mathit{o3}) + f_{\tau,3}(\mathit{prec}) + f_{\tau,4}(\mathit{temp}) + f_{\tau,5}(\mathit{vel}) + f_{\tau,6}(\mathit{racha}) \\
&\qquad\qquad + f_{\tau,7}(\mathit{pres\_max}) + f_{\tau,8}(\mathit{pres\_min}) + f_{\tau,9}(\mathit{traffic})
\end{aligned}\end{equation*}
Here, $f_{\tau,j}=f_{\tau,j,\pen}+f_{\tau,j,\unpen}$ has been decomposed into respective linear and nonlinear parts for each covariate (subject to selection) and $\beta_{\tau,k}$, $k=0,\ldots,3$ are the overall intercept and year-specific coefficients (not subject to selection). After inspecting Figures \ref{fig:prelim:no2:o3} and \ref{fig:prelim:no2:traffic}, we noted that NO$_2$ distributions differ depending on the predictor values, and also show a non-Gaussian behavior. We also underlined that the variance of NO$_2$ increases with increasing traffic and that the predictor O$_3$ has a clear nonlinear effect on NO$_2$. We note that the apparent linearity we observed between traffic and NO$_2$ in Figure \ref{fig:prelim:no2:traffic} is kind of spurious due to the increasing variability of the response, which makes the linearity assumption too restrictive. This will be reinforced later in this section.

\begin{figure}[htbp]
	\begin{center}
		\centering\includegraphics*[width=0.9\textwidth,angle=0,height=0.33\textheight]{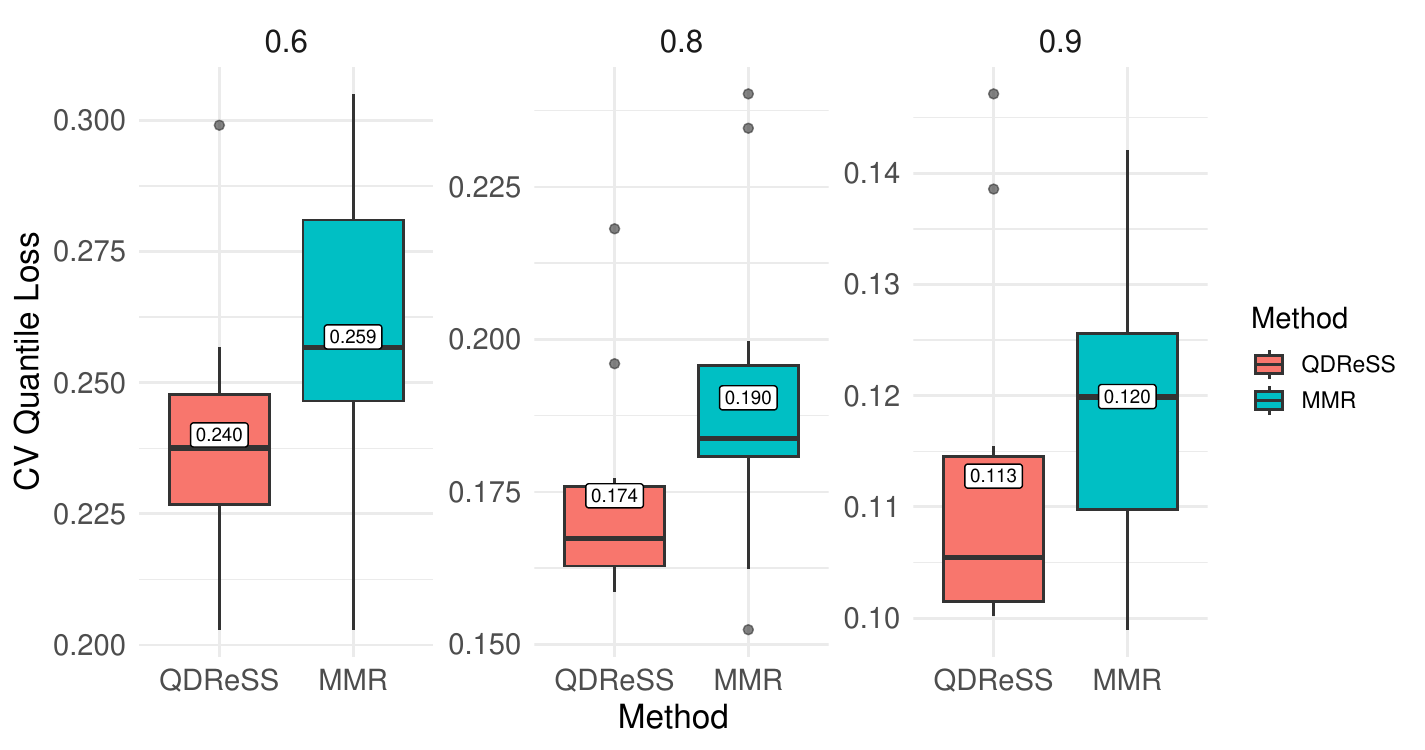}
	\end{center}
\caption{Boxplots of the out-of-sample quantile loss from 10-fold cross-validation for each method and quantile level $\tau$. The boxes represent the interquartile range across folds, with the median marked by the horizontal line. The mean quantile losses are reported as text.}
\label{fig:cv_loss}
\end{figure}

\paragraph*{Predictive performance}
To assess the out-of-sample predictive performance of \modnamesp against MMR, we conducted a 10-fold cross-validation study. The data were randomly partitioned into ten approximately equal folds, where nine folds are then used for estimation and the remaining fold for prediction each, respectively. Predictive accuracy is evaluated using the quantile loss function at each quantile level $\tau$. Specifically, for a hold-out observation $y_i$ and its predicted $\tau$-quantile $\hat{y}_{i,\tau}=\hat\eta_{i,\tau}$ from \eqref{eq:mod}, the quantile loss is defined as
$\text{QL}_{i,\tau}(y_i\mid x_i) = \rho_{\tau}(y_i- \hat{y}_{i,\tau}),$ where $\rho_\tau$ is the check function at \eqref{eq:awad}. The predicted quantile $\hat\eta_{i,\tau}$ is obtained via posterior mean estimates trained on $\approx 90\%$ of the data for each of the ten CV folds.  Figure \ref{fig:cv_loss} shows boxplots of the quantile loss together with its median (black horizontal line)  the first and third quartiles, and its mean $n^{-1}\sum_{i=1}^n\text{QL}_{i,\tau}(y_i\mid x_i)$ (value indicated as text) to summarize variability across CV folds. For all thresholds, our proposed method \modnamesp consistently achieves lower quantile loss than the competing approach MMR, indicating superior out-of-sample predictive performance. In light of these results, we proceed with a detailed analysis of the full dataset using the \modname.  MCMC diagnostics supporting  posterior inference are reported in the Supplementary Material D.

\paragraph*{Results}
\begin{table}[htbp]
\centering\renewcommand\arraystretch{1.00}
\begin{tabular}{c|cccc}
  \hline\hline
 Covariate & $f_{\tau,j,\unpen}/f_{\tau,j,\pen}$ & $\tau=0.6$ & $\tau=0.8$ & $\tau=0.9$ \\ 
  \hline
{\emph{co}} & $f_{\tau,1,\unpen,}$ & \textbf{1.000} & \textbf{0.785} & 0.041 \\ 
  {\emph{co}} & $f_{\tau,1,\pen}$ & \textbf{1.000} & 0.076 & 0.024 \\ 
  \hline
  {\emph{o3}} & $f_{\tau,2,\unpen}$ & \textbf{1.000} & \textbf{1.000} & \textbf{1.000} \\ 
  {\emph{o3}} & $f_{\tau,2,\pen}$ & \textbf{1.000} & \textbf{1.000} & \textbf{1.000} \\ 
  \hline
  {\emph{prec}} & $f_{\tau,3,\unpen}$ & 0.052 & 0.085 & \textbf{0.817} \\ 
  {\emph{prec}} & $f_{\tau,3\pen}$ & \textbf{1.000} & \textbf{1.000} & \textbf{1.000} \\ 
  \hline
  {\emph{temp}} & $f_{\tau,4,\unpen}$ & \textbf{1.000} & \textbf{1.000} & \textbf{1.000} \\ 
  {\emph{temp}} & $f_{\tau,4,\pen}$ & \textbf{1.000} & \textbf{1.000} & \textbf{1.000} \\ 
  \hline
  {\emph{vel}} & $f_{\tau,5,\unpen}$ & \textbf{0.990} & \textbf{1.000} & \textbf{1.000} \\ 
  {\emph{vel}} & $f_{\tau,5,\pen}$ & 0.007 & \textbf{1.000} & \textbf{1.000} \\ 
  \hline
  {\emph{racha}} & $f_{\tau,6,\unpen}$ & \textbf{1.000} & \textbf{1.000} & \textbf{1.000} \\ 
  {\emph{racha}} & $f_{\tau,6,\pen}$ & \textbf{1.000} & \textbf{1.000} & \textbf{1.000} \\ 
  \hline
  {\emph{pres\_max}} & $f_{\tau,7,\unpen}$ & 0.031 & 0.346 & 0.027 \\ 
  {\emph{pres\_max}} & $f_{\tau,7,\pen}$ & \textbf{0.559} & 0.187 & 0.095 \\ 
  \hline
  {\emph{pres\_min}} & $f_{\tau,8,\unpen}$ & \textbf{1.000} & \textbf{0.675} & \textbf{0.999} \\ 
  {\emph{pres\_min}} & $f_{\tau,8,\pen}$ & \textbf{1.000} & \textbf{0.696} & \textbf{0.943} \\ 
  \hline
  {\emph{traffic}} & $f_{\tau,9,\unpen}$ & \textbf{1.000} & \textbf{1.000} & \textbf{1.000} \\ 
  {\emph{traffic}} & $f_{\tau,9,\pen}$ & \textbf{1.000} & \textbf{1.000} & \textbf{0.628} \\ 
   \hline \hline
\end{tabular}
\caption{{Posterior mean inclusion probabilities of $f_{\tau,j,\unpen}$ (linear parts) and $f_{\tau,j,\pen}$ (nonlinear parts) for $\tau\in\lbrace 0.6,0.8,0.9\rbrace$ (across columns 3--5). We say an effect part should be included in the model if $\widehat p_{\tau,j}\geq 0.5$.}}\label{tab:inc}
\end{table}

\begin{figure}[htbp]
	\begin{center}
		\centering\includegraphics*[width=0.9\textwidth,angle=0]{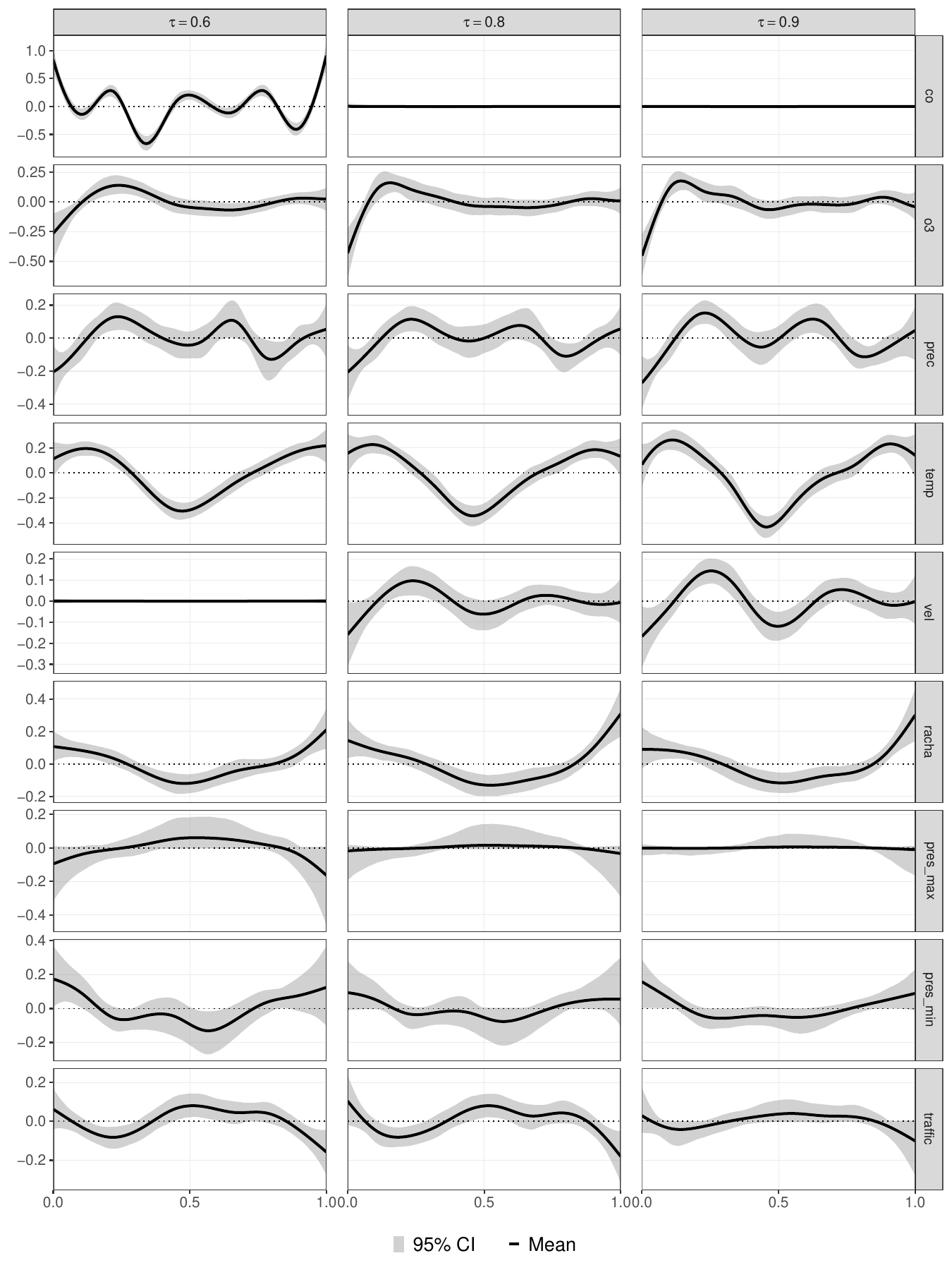}
	\end{center}
\caption{{Estimated posterior effects for $f_{\tau,j,\pen}$ (nonlinear parts) for $\tau\in\lbrace 0.6,0.8,0.9\rbrace$ (across columns 1--3) and for all nine covariates (row-wise). Shown are the posterior mean (solid lines) and 95\% pointwise credible intervals (shaded area).}}
	\label{fig:no2_nonlin}
\end{figure}

\begin{figure}[htbp]
	\begin{center}
		\centering\includegraphics*[width=0.9\textwidth,angle=0]{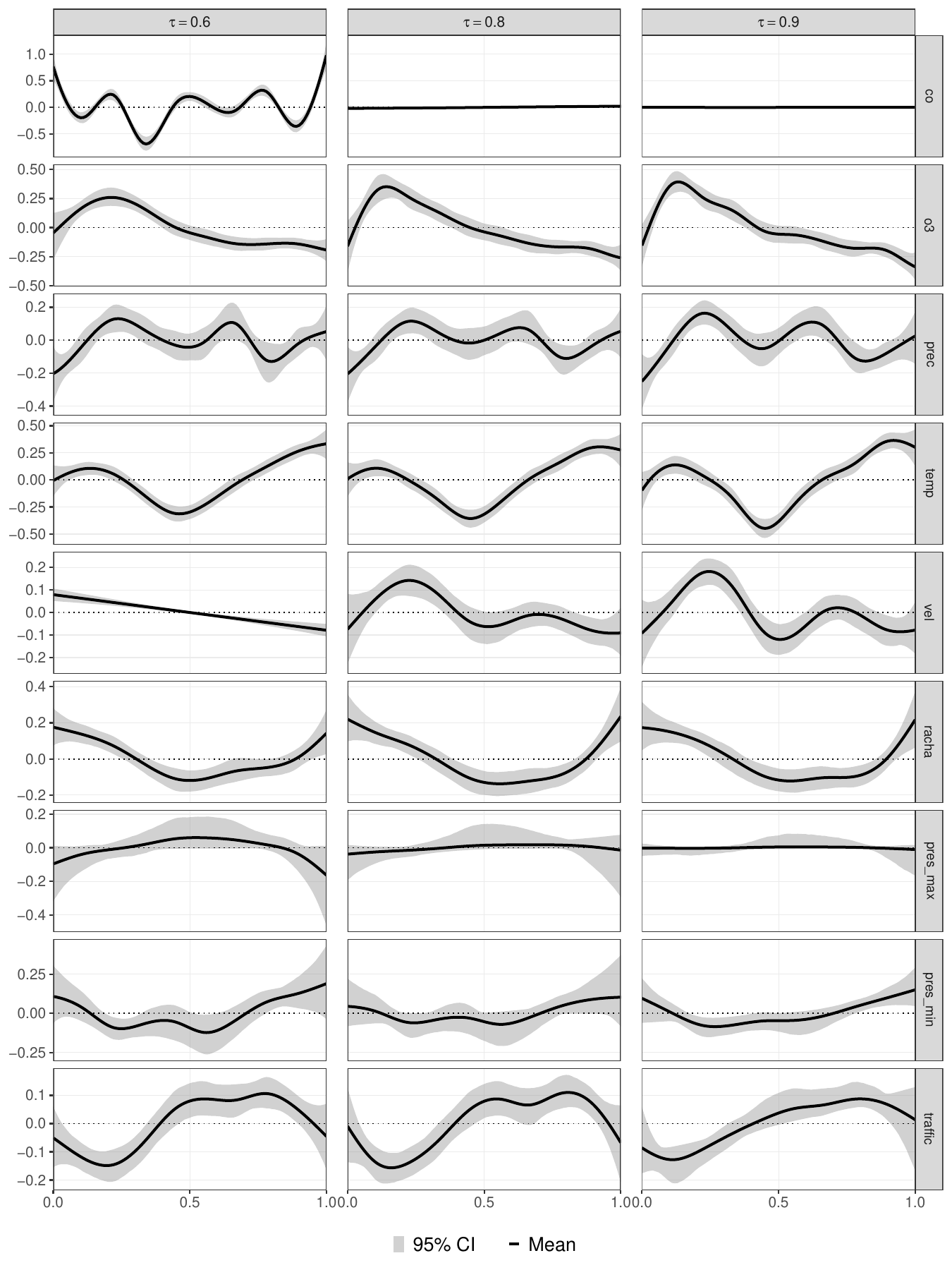}
	\end{center}
\caption{{Estimated posterior effects for $f_{\tau,j}=f_{\tau,j,\pen}+f_{\tau,j,\unpen}$ (linear parts+nonlinear parts) for $\tau\in\lbrace 0.6,0.8,0.9\rbrace$ (across columns 1--3) and for all nine covariates (row-wise). Shown are the posterior mean (solid lines) and 95\% pointwise credible intervals (shaded area).}}
	\label{fig:no2_both}
\end{figure}

 Table~\ref{tab:inc} shows posterior mean inclusion probabilities of $f_{\tau,j,\unpen}$ (linear parts) and $f_{\tau,j,\pen}$ (nonlinear parts) for $\tau\in\lbrace 0.6,0.8,0.9\rbrace$ (across columns 2--4). We {say that} an effect part should be included in the model if the corresponding posterior mean inclusion probability $\hat p_{\tau,j}\geq 0.5$ from \eqref{eq:pip} (in bold in Table~\ref{tab:inc}). In addition, Table~\ref{tab:no2_lin}, and Figures~\ref{fig:no2_nonlin} and~\ref{fig:no2_both} show estimated posterior effects for $\widehat f_{\tau,j, \unpen}$ (linear parts), $\widehat f_{\tau,j,\pen}$ (nonlinear parts) and $\widehat f_{\tau,j}=\widehat f_{\tau,j,\unpen}+\widehat f_{\tau,j,\pen}$ (linear parts+nonlinear parts) for $\tau\in\lbrace 0.6,0.8,0.9\rbrace$ (across columns 1--3) and for all nine covariates $j\in\lbrace 1,\ldots,9\rbrace$ (row-wise). Shown are the posterior mean (solid lines; from \eqref{eq:postmean}) and 95\% pointwise credible intervals (dashed lines; from \eqref{eq:ci}). For nonlinear components that are not selected, the credible intervals can be visually very narrow, reflecting strong posterior shrinkage of the corresponding effect part toward zero.

\begin{table}[htbp]
\centering
\begin{tabular}{l|ccc}
  \hline \hline
 Covariate & $\tau=0.6$ & $\tau=0.8$ & $\tau=0.9$ \\ 
  \hline
{\emph{co}} & 0.136 [0.102, 0.169] & 0.044 [-0.003, 0.081] & 0.002 [-0.006, 0.015] \\ 
  {\emph{o3}} & -0.435 [-0.489, -0.379] & -0.542 [-0.597, -0.486] & -0.597 [-0.644, -0.550] \\ 
  {\emph{prec}} & -0.002 [-0.018, 0.009] & -0.004 [-0.029, 0.008] & -0.042 [-0.075, 0.001] \\ 
  {\emph{temp}} & 0.234 [0.188, 0.280] & 0.291 [0.244, 0.337] & 0.323 [0.282, 0.364] \\ 
  {\emph{vel}} & -0.158 [-0.211, -0.111] & -0.170 [-0.216, -0.117] & -0.152 [-0.200, -0.105] \\ 
  {\emph{racha}} & -0.137 [-0.188, -0.086] & -0.149 [-0.206, -0.098] & -0.168 [-0.215, -0.119] \\ 
  {\emph{pres\_max}} & 0.001 [-0.016, 0.016] & 0.040 [-0.013, 0.148] & 0.001 [-0.016, 0.016] \\ 
  {\emph{pres\_min}} & 0.130 [0.085, 0.171] & 0.097 [-0.008, 0.179] & 0.123 [0.083, 0.165] \\ 
  {\emph{traffic}} & 0.226 [0.197, 0.258] & 0.227 [0.195, 0.258] & 0.229 [0.198, 0.261] \\ 
   \hline \hline
\end{tabular}
\caption{Estimated posterior effects for $f_{\tau,j,\unpen}$ (linear parts) for $\tau\in\lbrace 0.6,0.8,0.9\rbrace$ (across columns 2--4) and for all nine covariates (row-wise). Shown are the posterior mean and 95\% pointwise credible intervals (in brackets).}\label{tab:no2_lin}
\end{table}

We highlight the following results. CO ($\mathit{co}$) has a strong positive linear effect for $\tau=0.6$, but is negligible for $\tau=0.8,0.9$. This indicates that CO contributes to the NO$_2$ distribution in a linear form but only for lower thresholds. In contrast, O$_3$ ($\mathit{o3}$) is a good predictor with a clear nonlinear effect for all three quantiles with a notable decreasing effect. This inverse relationship between NO$_2$ and O$_3$ is expected as we discussed in Section \ref{subsec:motivation}. However, the nonlinear structure was not so apparent, and we are able to underpin it. 

Precipitation ($\mathit{prec}$) enters as a nonlinear effect, and varies its strength depending on the quantile. In particular, the effect and its nonlinearity increase with increasing quantile $\tau$. This combination of nonlinear behavior and increasing strength with larger quantiles provides a clear explanation of the effect of precipitation over NO$_2$. Average temperature ($\mathit{temp}$) enters as both  nonlinear and linear effects for $\tau=0.6,0.9$  but only as a nonlinear effect for $\tau=0.8,0.9$. Thus, the quantile has an effect on the dependence between  NO$_2$ and average temperature with the higher the temperature, the larger NO$_2$.

Average wind speed ($\mathit{vel}$) enters as a nonlinear effect for $\tau=0.8,0.9$ and linearly for $\tau=0.6$. The linear effect for small thresholds is inverse indicating that with an increasing wind speed we get decreasing NO$_2$ values. Importantly, for larger thresholds this linearity vanishes towards nonlinear effects. When coming to maximum wind gusts ($\mathit{racha}$), the effects are basically  nonlinear for all quantiles. Air pressure ($\mathit{pres\_min}$) is only relevant as a linear effect for $\tau=0.8$ measured by minimum air pressure, but negligible for all other quantiles. Maximum pressure ($\mathit{pres\_max}$) is however not selected.

Finally, average traffic flow ($\mathit{traffic}$)  enters as a nonlinear effect, with a stronger effect for $\tau=0.6,0.9$ and weaker effect for $\tau=0.8$. This indicates that strong traffic congestion{s} affect NO$_2$ in a complicated nonlinear fashion, but this also holds for lower thresholds, probably due to  cross-relationships among some of the covariates. Latent (unobserved variables) might also place a hidden effect here, and are difficult to account for.

Noting that modeling reality of air pollution is 
certainly a critical, while complicated environmental problem, we have detected some functional forms, some of them highly nonlinear, that underline the cross-relationships between a number of covariates and NO$_2$. Chemical reactions are playing a role and make things even more complicated. Our statistical approach has been able to clarify some of these complicated relations.

\section{Discussion and conclusion}\label{sec:conclusion}

We have proposed \modname, a Bayesian framework for effect selection in semi-parametric quantile regression models. Inspired by the recent work of \citet{KleCarKneLanWag2021} in the context of structured additive distributional regression~\citep{Kle2024}, \modnamesp employs an NBPSS prior on the scalar squared importance parameters associated with each effect  in the predictor. Posterior estimation is carried out using an efficient Gibbs sampler. Compared to the distributional models of~\cite{KleKneKlaLan2015}, where predictors are placed on the distributional parameters, \modnamesp can directly select certain effect types on conditional quantiles of a response, and decide whether relevant predictors affect the quantiles linearly, nonlinearly, both, or not at all. To enable consistent estimation of both effect parts, we suggest replacing the commonly employed MMR \citep[in e.g., ][]{KleKneKlaLan2015,KleCarKneLanWag2021} by  a novel DR basis expansion recently suggested by \citet{BacKle2024}. In contrast to the MMR, this parameterization enforces orthogonality between linear and nonlinear effects of a specific covariate. As a theoretical contribution, we have shown that the resulting linear and nonlinear effect components can be estimated consistently. This result provides a formal basis for interpreting and selecting the two effect parts on conditional quantiles separately. The DR parameterization also offers additional computational advantages over the MMR. Finally, we suggest fast prior elicitation by exploiting the scaling property of the (scaled) beta prime distribution as opposed to time-consuming numerical optimization that was previously proposed. 

While our framework is applicable to a wide range of applications where interest is in understanding the impacts of influential covariates on specific conditional quantiles of a dependent variable, our methodological developments are specifically motivated by one of the most pressing environmental health risks in Europe: air pollution. Many European cities regularly exceed NO$_2$ limits, and we hope to bring this issue to light by analyzing data obtained from a particular weather and pollution station in downtown Madrid.  We argue that our framework can more accurately capture the reality of air pollution by detecting complicated linear and nonlinear functional forms in combination with particular alarm thresholds.  The results of this study are easily extendable to many other cities worldwide, perhaps with appropriate adaptations and the use of additional covariates. Indeed, our statistical approach enables the study of quantile-specific covariate effects across general functional forms and facilitates the decision of whether an effect should be included linearly, nonlinearly, or not at all in the relevant threshold quantiles.  

We note at this point that this paper focuses on a single representative station in a large city  as Madrid. This choice reflects our primary objective of analyzing the relationships between multiple predictors and NO$_2$ concentrations to better understand the intrinsic mechanisms underlying air pollution, rather than focusing on prediction. These complex mechanisms are often overlooked or inadequately captured by previous statistical approaches.  We are also aware that there are other measuring stations spread throughout the city, where the spatial structure could be relevant if the focus were on imputing missing data or forecasting. We leave this important point as a direction for future research. 

From a methodological perspective, it is worth noting that estimating quantiles of interest simultaneously, rather than separately, can be useful to avoid quantile crossing in applications where a more dense grid of quantiles is of interest. Investigating robustness to the ALD assumption itself, including alternative loss-based posteriors, is a natural extension. Finally, other effect types, such as spatial covariates or individual specific random effects, can be incorporated in a straightforward manner.

\bibliographystyle{dcu}
\bibliography{bibliography}

\end{document}